\documentclass[longauth]{aa}

\usepackage{graphicx}
\usepackage{natbib}
\usepackage{scalerel}
\usepackage{xcolor}
\usepackage[normalem]{ulem}

\usepackage{txfonts}
\usepackage[pdfencoding=auto,psdextra]{hyperref}
\hypersetup{
    colorlinks=true,
   linkcolor=blue,
    filecolor=magenta,      
    urlcolor=blue,
    citecolor=blue,
    pdftitle={Euclid: Dwarf galaxies in the Q1 data release},
    pdfauthor={Marleau et al.}
}
\makeatletter
\renewcommand*\aa@pageof{, page \thepage{} of \pageref*{LastPage}}
\makeatother
\usepackage{lastpage}

\usepackage{orcidlink}
\newcommand{\orcid}[1]{\orcidlink{#1}}

\usepackage[utf8]{inputenc}

\usepackage[switch, modulo]{lineno}

\usepackage{euclid}
\usepackage{rotating}
\usepackage{pdflscape}

\newcommand{\magarc}{mag\,arcsec$^{-2}$}

\begin{document}

\title{\Euclid: Quick Data Release (Q1) -- A census of dwarf galaxies across a range of distances and environments\thanks{This paper is published on behalf of the Euclid Consortium.}}    

%List of EC publications: https://pubs.euclid-ec.org/
%Author list: https://pubs.euclid-ec.org/?publication-id=PUB246ae73a958bea88
%Upload: https://pubs.euclid-ec.org/upload.php?publication-id=PUB246ae73a958bea88

%%%% Version Friday 14th of March 2025 10:09:12 AM UT												
%%%% Please do not edit the author list -- contact ECEB Bureau for changes
%\newcommand{\orcid}[1]{} %% if already defined in aa.cls: comment, or use renewcommand			   
\author{F.~R.~Marleau\orcid{0000-0002-1442-2947}\thanks{\email{francine.marleau@uibk.ac.at}}\inst{\ref{aff1}}
\and R.~Habas\orcid{0000-0002-4033-3841}\inst{\ref{aff2}}
\and D.~Carollo\orcid{0000-0002-0005-5787}\inst{\ref{aff3}}
\and C.~Tortora\orcid{0000-0001-7958-6531}\inst{\ref{aff4}}
\and P.-A.~Duc\orcid{0000-0003-3343-6284}\inst{\ref{aff5}}
\and E.~Sola\orcid{0000-0002-2814-3578}\inst{\ref{aff6}}
\and T.~Saifollahi\orcid{0000-0002-9554-7660}\inst{\ref{aff5}}
\and M.~F\"ugenschuh\orcid{0000-0002-1216-6003}\inst{\ref{aff1}}
\and M.~Walmsley\orcid{0000-0002-6408-4181}\inst{\ref{aff7},\ref{aff8}}
\and R.~Z\"oller\orcid{0000-0002-0938-5686}\inst{\ref{aff9},\ref{aff10}}
\and A.~Ferr\'e-Mateu\orcid{0000-0002-6411-220X}\inst{\ref{aff11},\ref{aff12}}
\and M.~Cantiello\orcid{0000-0003-2072-384X}\inst{\ref{aff2}}
\and M.~Urbano\orcid{0000-0001-5640-0650}\inst{\ref{aff5}}
\and E.~Saremi\orcid{0000-0002-5075-1764}\inst{\ref{aff13}}
\and R.~Ragusa\inst{\ref{aff4}}
\and R.~Laureijs\inst{\ref{aff14},\ref{aff15}}
\and M.~Hilker\orcid{0000-0002-2363-5522}\inst{\ref{aff16}}
\and O.~M\"uller\orcid{0000-0003-4552-9808}\inst{\ref{aff17},\ref{aff6},\ref{aff18}}
\and M.~Poulain\orcid{0000-0002-7664-4510}\inst{\ref{aff19}}
\and R.~F.~Peletier\orcid{0000-0001-7621-947X}\inst{\ref{aff14}}
\and S.~J.~Sprenger\inst{\ref{aff1}}
\and O.~Marchal\orcid{0000-0001-7461-8928}\inst{\ref{aff5}}
\and N.~Aghanim\orcid{0000-0002-6688-8992}\inst{\ref{aff20}}
\and B.~Altieri\orcid{0000-0003-3936-0284}\inst{\ref{aff21}}
\and A.~Amara\inst{\ref{aff22}}
\and S.~Andreon\orcid{0000-0002-2041-8784}\inst{\ref{aff23}}
\and N.~Auricchio\orcid{0000-0003-4444-8651}\inst{\ref{aff24}}
\and H.~Aussel\orcid{0000-0002-1371-5705}\inst{\ref{aff25}}
\and C.~Baccigalupi\orcid{0000-0002-8211-1630}\inst{\ref{aff26},\ref{aff3},\ref{aff27},\ref{aff28}}
\and M.~Baldi\orcid{0000-0003-4145-1943}\inst{\ref{aff29},\ref{aff24},\ref{aff30}}
\and A.~Balestra\orcid{0000-0002-6967-261X}\inst{\ref{aff31}}
\and S.~Bardelli\orcid{0000-0002-8900-0298}\inst{\ref{aff24}}
\and A.~Basset\inst{\ref{aff32}}
\and P.~Battaglia\orcid{0000-0002-7337-5909}\inst{\ref{aff24}}
\and R.~Bender\orcid{0000-0001-7179-0626}\inst{\ref{aff10},\ref{aff9}}
\and A.~Biviano\orcid{0000-0002-0857-0732}\inst{\ref{aff3},\ref{aff26}}
\and A.~Bonchi\orcid{0000-0002-2667-5482}\inst{\ref{aff33}}
\and D.~Bonino\orcid{0000-0002-3336-9977}\inst{\ref{aff34}}
\and E.~Branchini\orcid{0000-0002-0808-6908}\inst{\ref{aff35},\ref{aff36},\ref{aff23}}
\and M.~Brescia\orcid{0000-0001-9506-5680}\inst{\ref{aff37},\ref{aff4}}
\and J.~Brinchmann\orcid{0000-0003-4359-8797}\inst{\ref{aff38},\ref{aff39}}
\and S.~Camera\orcid{0000-0003-3399-3574}\inst{\ref{aff40},\ref{aff41},\ref{aff34}}
\and G.~Ca\~nas-Herrera\orcid{0000-0003-2796-2149}\inst{\ref{aff15},\ref{aff42},\ref{aff43}}
\and V.~Capobianco\orcid{0000-0002-3309-7692}\inst{\ref{aff34}}
\and C.~Carbone\orcid{0000-0003-0125-3563}\inst{\ref{aff44}}
\and J.~Carretero\orcid{0000-0002-3130-0204}\inst{\ref{aff45},\ref{aff46}}
\and S.~Casas\orcid{0000-0002-4751-5138}\inst{\ref{aff47}}
\and M.~Castellano\orcid{0000-0001-9875-8263}\inst{\ref{aff48}}
\and G.~Castignani\orcid{0000-0001-6831-0687}\inst{\ref{aff24}}
\and S.~Cavuoti\orcid{0000-0002-3787-4196}\inst{\ref{aff4},\ref{aff49}}
\and K.~C.~Chambers\orcid{0000-0001-6965-7789}\inst{\ref{aff50}}
\and A.~Cimatti\inst{\ref{aff51}}
\and C.~Colodro-Conde\inst{\ref{aff11}}
\and G.~Congedo\orcid{0000-0003-2508-0046}\inst{\ref{aff52}}
\and C.~J.~Conselice\orcid{0000-0003-1949-7638}\inst{\ref{aff8}}
\and L.~Conversi\orcid{0000-0002-6710-8476}\inst{\ref{aff53},\ref{aff21}}
\and Y.~Copin\orcid{0000-0002-5317-7518}\inst{\ref{aff54}}
\and L.~Corcione\orcid{0000-0002-6497-5881}\inst{\ref{aff34}}
\and F.~Courbin\orcid{0000-0003-0758-6510}\inst{\ref{aff55},\ref{aff56}}
\and H.~M.~Courtois\orcid{0000-0003-0509-1776}\inst{\ref{aff57}}
\and M.~Cropper\orcid{0000-0003-4571-9468}\inst{\ref{aff58}}
\and J.-C.~Cuillandre\orcid{0000-0002-3263-8645}\inst{\ref{aff25}}
\and A.~Da~Silva\orcid{0000-0002-6385-1609}\inst{\ref{aff59},\ref{aff60}}
\and H.~Degaudenzi\orcid{0000-0002-5887-6799}\inst{\ref{aff61}}
\and G.~De~Lucia\orcid{0000-0002-6220-9104}\inst{\ref{aff3}}
\and A.~M.~Di~Giorgio\orcid{0000-0002-4767-2360}\inst{\ref{aff62}}
\and C.~Dolding\orcid{0009-0003-7199-6108}\inst{\ref{aff58}}
\and H.~Dole\orcid{0000-0002-9767-3839}\inst{\ref{aff20}}
\and F.~Dubath\orcid{0000-0002-6533-2810}\inst{\ref{aff61}}
\and X.~Dupac\inst{\ref{aff21}}
\and S.~Dusini\orcid{0000-0002-1128-0664}\inst{\ref{aff63}}
\and S.~Escoffier\orcid{0000-0002-2847-7498}\inst{\ref{aff64}}
\and M.~Fabricius\orcid{0000-0002-7025-6058}\inst{\ref{aff10},\ref{aff9}}
\and M.~Farina\orcid{0000-0002-3089-7846}\inst{\ref{aff62}}
\and F.~Faustini\orcid{0000-0001-6274-5145}\inst{\ref{aff33},\ref{aff48}}
\and S.~Ferriol\inst{\ref{aff54}}
\and P.~Fosalba\orcid{0000-0002-1510-5214}\inst{\ref{aff65},\ref{aff66}}
\and S.~Fotopoulou\orcid{0000-0002-9686-254X}\inst{\ref{aff67}}
\and M.~Frailis\orcid{0000-0002-7400-2135}\inst{\ref{aff3}}
\and E.~Franceschi\orcid{0000-0002-0585-6591}\inst{\ref{aff24}}
\and P.~Franzetti\inst{\ref{aff44}}
\and M.~Fumana\orcid{0000-0001-6787-5950}\inst{\ref{aff44}}
\and S.~Galeotta\orcid{0000-0002-3748-5115}\inst{\ref{aff3}}
\and K.~George\orcid{0000-0002-1734-8455}\inst{\ref{aff9}}
\and B.~Gillis\orcid{0000-0002-4478-1270}\inst{\ref{aff52}}
\and C.~Giocoli\orcid{0000-0002-9590-7961}\inst{\ref{aff24},\ref{aff30}}
\and B.~R.~Granett\orcid{0000-0003-2694-9284}\inst{\ref{aff23}}
\and A.~Grazian\orcid{0000-0002-5688-0663}\inst{\ref{aff31}}
\and F.~Grupp\inst{\ref{aff10},\ref{aff9}}
\and S.~Gwyn\orcid{0000-0001-8221-8406}\inst{\ref{aff68}}
\and S.~V.~H.~Haugan\orcid{0000-0001-9648-7260}\inst{\ref{aff69}}
\and J.~Hoar\inst{\ref{aff21}}
\and H.~Hoekstra\orcid{0000-0002-0641-3231}\inst{\ref{aff43}}
\and W.~Holmes\inst{\ref{aff70}}
\and F.~Hormuth\inst{\ref{aff71}}
\and A.~Hornstrup\orcid{0000-0002-3363-0936}\inst{\ref{aff72},\ref{aff73}}
\and P.~Hudelot\inst{\ref{aff74}}
\and K.~Jahnke\orcid{0000-0003-3804-2137}\inst{\ref{aff75}}
\and M.~Jhabvala\inst{\ref{aff76}}
\and B.~Joachimi\orcid{0000-0001-7494-1303}\inst{\ref{aff77}}
\and E.~Keih\"anen\orcid{0000-0003-1804-7715}\inst{\ref{aff78}}
\and S.~Kermiche\orcid{0000-0002-0302-5735}\inst{\ref{aff64}}
\and A.~Kiessling\orcid{0000-0002-2590-1273}\inst{\ref{aff70}}
\and B.~Kubik\orcid{0009-0006-5823-4880}\inst{\ref{aff54}}
\and M.~K\"ummel\orcid{0000-0003-2791-2117}\inst{\ref{aff9}}
\and M.~Kunz\orcid{0000-0002-3052-7394}\inst{\ref{aff79}}
\and H.~Kurki-Suonio\orcid{0000-0002-4618-3063}\inst{\ref{aff80},\ref{aff81}}
\and O.~Lahav\orcid{0000-0002-1134-9035}\inst{\ref{aff77}}
\and Q.~Le~Boulc'h\inst{\ref{aff82}}
\and A.~M.~C.~Le~Brun\orcid{0000-0002-0936-4594}\inst{\ref{aff83}}
\and D.~Le~Mignant\orcid{0000-0002-5339-5515}\inst{\ref{aff84}}
\and S.~Ligori\orcid{0000-0003-4172-4606}\inst{\ref{aff34}}
\and P.~B.~Lilje\orcid{0000-0003-4324-7794}\inst{\ref{aff69}}
\and V.~Lindholm\orcid{0000-0003-2317-5471}\inst{\ref{aff80},\ref{aff81}}
\and I.~Lloro\orcid{0000-0001-5966-1434}\inst{\ref{aff85}}
\and G.~Mainetti\orcid{0000-0003-2384-2377}\inst{\ref{aff82}}
\and D.~Maino\inst{\ref{aff86},\ref{aff44},\ref{aff87}}
\and E.~Maiorano\orcid{0000-0003-2593-4355}\inst{\ref{aff24}}
\and O.~Mansutti\orcid{0000-0001-5758-4658}\inst{\ref{aff3}}
\and S.~Marcin\inst{\ref{aff88}}
\and O.~Marggraf\orcid{0000-0001-7242-3852}\inst{\ref{aff89}}
\and M.~Martinelli\orcid{0000-0002-6943-7732}\inst{\ref{aff48},\ref{aff90}}
\and N.~Martinet\orcid{0000-0003-2786-7790}\inst{\ref{aff84}}
\and F.~Marulli\orcid{0000-0002-8850-0303}\inst{\ref{aff91},\ref{aff24},\ref{aff30}}
\and R.~Massey\orcid{0000-0002-6085-3780}\inst{\ref{aff92}}
\and S.~Maurogordato\inst{\ref{aff93}}
\and H.~J.~McCracken\orcid{0000-0002-9489-7765}\inst{\ref{aff74}}
\and E.~Medinaceli\orcid{0000-0002-4040-7783}\inst{\ref{aff24}}
\and S.~Mei\orcid{0000-0002-2849-559X}\inst{\ref{aff94},\ref{aff95}}
\and M.~Melchior\inst{\ref{aff88}}
\and Y.~Mellier\inst{\ref{aff96},\ref{aff74}}
\and M.~Meneghetti\orcid{0000-0003-1225-7084}\inst{\ref{aff24},\ref{aff30}}
\and E.~Merlin\orcid{0000-0001-6870-8900}\inst{\ref{aff48}}
\and G.~Meylan\inst{\ref{aff17}}
\and A.~Mora\orcid{0000-0002-1922-8529}\inst{\ref{aff97}}
\and M.~Moresco\orcid{0000-0002-7616-7136}\inst{\ref{aff91},\ref{aff24}}
\and L.~Moscardini\orcid{0000-0002-3473-6716}\inst{\ref{aff91},\ref{aff24},\ref{aff30}}
\and R.~Nakajima\orcid{0009-0009-1213-7040}\inst{\ref{aff89}}
\and C.~Neissner\orcid{0000-0001-8524-4968}\inst{\ref{aff98},\ref{aff46}}
\and S.-M.~Niemi\inst{\ref{aff15}}
\and J.~W.~Nightingale\orcid{0000-0002-8987-7401}\inst{\ref{aff99}}
\and C.~Padilla\orcid{0000-0001-7951-0166}\inst{\ref{aff98}}
\and S.~Paltani\orcid{0000-0002-8108-9179}\inst{\ref{aff61}}
\and F.~Pasian\orcid{0000-0002-4869-3227}\inst{\ref{aff3}}
\and K.~Pedersen\inst{\ref{aff100}}
\and W.~J.~Percival\orcid{0000-0002-0644-5727}\inst{\ref{aff101},\ref{aff102},\ref{aff103}}
\and V.~Pettorino\inst{\ref{aff15}}
\and S.~Pires\orcid{0000-0002-0249-2104}\inst{\ref{aff25}}
\and G.~Polenta\orcid{0000-0003-4067-9196}\inst{\ref{aff33}}
\and M.~Poncet\inst{\ref{aff32}}
\and L.~A.~Popa\inst{\ref{aff104}}
\and L.~Pozzetti\orcid{0000-0001-7085-0412}\inst{\ref{aff24}}
\and F.~Raison\orcid{0000-0002-7819-6918}\inst{\ref{aff10}}
\and R.~Rebolo\inst{\ref{aff105},\ref{aff12},\ref{aff11}}
\and A.~Renzi\orcid{0000-0001-9856-1970}\inst{\ref{aff106},\ref{aff63}}
\and J.~Rhodes\orcid{0000-0002-4485-8549}\inst{\ref{aff70}}
\and G.~Riccio\inst{\ref{aff4}}
\and E.~Romelli\orcid{0000-0003-3069-9222}\inst{\ref{aff3}}
\and M.~Roncarelli\orcid{0000-0001-9587-7822}\inst{\ref{aff24}}
\and E.~Rossetti\orcid{0000-0003-0238-4047}\inst{\ref{aff29}}
\and B.~Rusholme\orcid{0000-0001-7648-4142}\inst{\ref{aff107}}
\and R.~Saglia\orcid{0000-0003-0378-7032}\inst{\ref{aff9},\ref{aff10}}
\and Z.~Sakr\orcid{0000-0002-4823-3757}\inst{\ref{aff108},\ref{aff109},\ref{aff110}}
\and A.~G.~S\'anchez\orcid{0000-0003-1198-831X}\inst{\ref{aff10}}
\and D.~Sapone\orcid{0000-0001-7089-4503}\inst{\ref{aff111}}
\and B.~Sartoris\orcid{0000-0003-1337-5269}\inst{\ref{aff9},\ref{aff3}}
\and M.~Sauvage\orcid{0000-0002-0809-2574}\inst{\ref{aff25}}
\and J.~A.~Schewtschenko\orcid{0000-0002-4913-6393}\inst{\ref{aff52}}
\and M.~Schirmer\orcid{0000-0003-2568-9994}\inst{\ref{aff75}}
\and P.~Schneider\orcid{0000-0001-8561-2679}\inst{\ref{aff89}}
\and M.~Scodeggio\inst{\ref{aff44}}
\and A.~Secroun\orcid{0000-0003-0505-3710}\inst{\ref{aff64}}
\and G.~Seidel\orcid{0000-0003-2907-353X}\inst{\ref{aff75}}
\and M.~Seiffert\orcid{0000-0002-7536-9393}\inst{\ref{aff70}}
\and S.~Serrano\orcid{0000-0002-0211-2861}\inst{\ref{aff65},\ref{aff112},\ref{aff66}}
\and P.~Simon\inst{\ref{aff89}}
\and C.~Sirignano\orcid{0000-0002-0995-7146}\inst{\ref{aff106},\ref{aff63}}
\and G.~Sirri\orcid{0000-0003-2626-2853}\inst{\ref{aff30}}
\and J.~Skottfelt\orcid{0000-0003-1310-8283}\inst{\ref{aff113}}
\and L.~Stanco\orcid{0000-0002-9706-5104}\inst{\ref{aff63}}
\and J.~Steinwagner\orcid{0000-0001-7443-1047}\inst{\ref{aff10}}
\and P.~Tallada-Cresp\'{i}\orcid{0000-0002-1336-8328}\inst{\ref{aff45},\ref{aff46}}
\and D.~Tavagnacco\orcid{0000-0001-7475-9894}\inst{\ref{aff3}}
\and A.~N.~Taylor\inst{\ref{aff52}}
\and H.~I.~Teplitz\orcid{0000-0002-7064-5424}\inst{\ref{aff114}}
\and I.~Tereno\inst{\ref{aff59},\ref{aff115}}
\and S.~Toft\orcid{0000-0003-3631-7176}\inst{\ref{aff116},\ref{aff117}}
\and R.~Toledo-Moreo\orcid{0000-0002-2997-4859}\inst{\ref{aff118}}
\and F.~Torradeflot\orcid{0000-0003-1160-1517}\inst{\ref{aff46},\ref{aff45}}
\and I.~Tutusaus\orcid{0000-0002-3199-0399}\inst{\ref{aff109}}
\and L.~Valenziano\orcid{0000-0002-1170-0104}\inst{\ref{aff24},\ref{aff119}}
\and J.~Valiviita\orcid{0000-0001-6225-3693}\inst{\ref{aff80},\ref{aff81}}
\and T.~Vassallo\orcid{0000-0001-6512-6358}\inst{\ref{aff9},\ref{aff3}}
\and G.~Verdoes~Kleijn\orcid{0000-0001-5803-2580}\inst{\ref{aff14}}
\and A.~Veropalumbo\orcid{0000-0003-2387-1194}\inst{\ref{aff23},\ref{aff36},\ref{aff35}}
\and Y.~Wang\orcid{0000-0002-4749-2984}\inst{\ref{aff114}}
\and J.~Weller\orcid{0000-0002-8282-2010}\inst{\ref{aff9},\ref{aff10}}
\and A.~Zacchei\orcid{0000-0003-0396-1192}\inst{\ref{aff3},\ref{aff26}}
\and G.~Zamorani\orcid{0000-0002-2318-301X}\inst{\ref{aff24}}
\and F.~M.~Zerbi\inst{\ref{aff23}}
\and E.~Zucca\orcid{0000-0002-5845-8132}\inst{\ref{aff24}}
\and M.~Bolzonella\orcid{0000-0003-3278-4607}\inst{\ref{aff24}}
\and C.~Burigana\orcid{0000-0002-3005-5796}\inst{\ref{aff120},\ref{aff119}}
\and R.~Cabanac\orcid{0000-0001-6679-2600}\inst{\ref{aff109}}
\and L.~Gabarra\orcid{0000-0002-8486-8856}\inst{\ref{aff121}}
\and M.~Huertas-Company\orcid{0000-0002-1416-8483}\inst{\ref{aff11},\ref{aff122},\ref{aff123},\ref{aff124}}
\and V.~Scottez\inst{\ref{aff96},\ref{aff125}}
\and D.~Scott\orcid{0000-0002-6878-9840}\inst{\ref{aff126}}}
										   
%%%% please do not edit the affiliation list -- contact ECEB Bureau for changes
\institute{Universit\"at Innsbruck, Institut f\"ur Astro- und Teilchenphysik, Technikerstr. 25/8, 6020 Innsbruck, Austria\label{aff1}
\and
INAF - Osservatorio Astronomico d'Abruzzo, Via Maggini, 64100, Teramo, Italy\label{aff2}
\and
INAF-Osservatorio Astronomico di Trieste, Via G. B. Tiepolo 11, 34143 Trieste, Italy\label{aff3}
\and
INAF-Osservatorio Astronomico di Capodimonte, Via Moiariello 16, 80131 Napoli, Italy\label{aff4}
\and
Universit\'e de Strasbourg, CNRS, Observatoire astronomique de Strasbourg, UMR 7550, 67000 Strasbourg, France\label{aff5}
\and
Institute of Astronomy, University of Cambridge, Madingley Road, Cambridge CB3 0HA, UK\label{aff6}
\and
David A. Dunlap Department of Astronomy \& Astrophysics, University of Toronto, 50 St George Street, Toronto, Ontario M5S 3H4, Canada\label{aff7}
\and
Jodrell Bank Centre for Astrophysics, Department of Physics and Astronomy, University of Manchester, Oxford Road, Manchester M13 9PL, UK\label{aff8}
\and
Universit\"ats-Sternwarte M\"unchen, Fakult\"at f\"ur Physik, Ludwig-Maximilians-Universit\"at M\"unchen, Scheinerstrasse 1, 81679 M\"unchen, Germany\label{aff9}
\and
Max Planck Institute for Extraterrestrial Physics, Giessenbachstr. 1, 85748 Garching, Germany\label{aff10}
\and
Instituto de Astrof\'{\i}sica de Canarias, V\'{\i}a L\'actea, 38205 La Laguna, Tenerife, Spain\label{aff11}
\and
Universidad de La Laguna, Departamento de Astrof\'{\i}sica, 38206 La Laguna, Tenerife, Spain\label{aff12}
\and
School of Physics \& Astronomy, University of Southampton, Highfield Campus, Southampton SO17 1BJ, UK\label{aff13}
\and
Kapteyn Astronomical Institute, University of Groningen, PO Box 800, 9700 AV Groningen, The Netherlands\label{aff14}
\and
European Space Agency/ESTEC, Keplerlaan 1, 2201 AZ Noordwijk, The Netherlands\label{aff15}
\and
European Southern Observatory, Karl-Schwarzschild-Str.~2, 85748 Garching, Germany\label{aff16}
\and
Institute of Physics, Laboratory of Astrophysics, Ecole Polytechnique F\'ed\'erale de Lausanne (EPFL), Observatoire de Sauverny, 1290 Versoix, Switzerland\label{aff17}
\and
Visiting Fellow, Clare Hall, University of Cambridge, Cambridge, UK\label{aff18}
\and
Space physics and astronomy research unit, University of Oulu, Pentti Kaiteran katu 1, FI-90014 Oulu, Finland\label{aff19}
\and
Universit\'e Paris-Saclay, CNRS, Institut d'astrophysique spatiale, 91405, Orsay, France\label{aff20}
\and
ESAC/ESA, Camino Bajo del Castillo, s/n., Urb. Villafranca del Castillo, 28692 Villanueva de la Ca\~nada, Madrid, Spain\label{aff21}
\and
School of Mathematics and Physics, University of Surrey, Guildford, Surrey, GU2 7XH, UK\label{aff22}
\and
INAF-Osservatorio Astronomico di Brera, Via Brera 28, 20122 Milano, Italy\label{aff23}
\and
INAF-Osservatorio di Astrofisica e Scienza dello Spazio di Bologna, Via Piero Gobetti 93/3, 40129 Bologna, Italy\label{aff24}
\and
Universit\'e Paris-Saclay, Universit\'e Paris Cit\'e, CEA, CNRS, AIM, 91191, Gif-sur-Yvette, France\label{aff25}
\and
IFPU, Institute for Fundamental Physics of the Universe, via Beirut 2, 34151 Trieste, Italy\label{aff26}
\and
INFN, Sezione di Trieste, Via Valerio 2, 34127 Trieste TS, Italy\label{aff27}
\and
SISSA, International School for Advanced Studies, Via Bonomea 265, 34136 Trieste TS, Italy\label{aff28}
\and
Dipartimento di Fisica e Astronomia, Universit\`a di Bologna, Via Gobetti 93/2, 40129 Bologna, Italy\label{aff29}
\and
INFN-Sezione di Bologna, Viale Berti Pichat 6/2, 40127 Bologna, Italy\label{aff30}
\and
INAF-Osservatorio Astronomico di Padova, Via dell'Osservatorio 5, 35122 Padova, Italy\label{aff31}
\and
Centre National d'Etudes Spatiales -- Centre spatial de Toulouse, 18 avenue Edouard Belin, 31401 Toulouse Cedex 9, France\label{aff32}
\and
Space Science Data Center, Italian Space Agency, via del Politecnico snc, 00133 Roma, Italy\label{aff33}
\and
INAF-Osservatorio Astrofisico di Torino, Via Osservatorio 20, 10025 Pino Torinese (TO), Italy\label{aff34}
\and
Dipartimento di Fisica, Universit\`a di Genova, Via Dodecaneso 33, 16146, Genova, Italy\label{aff35}
\and
INFN-Sezione di Genova, Via Dodecaneso 33, 16146, Genova, Italy\label{aff36}
\and
Department of Physics "E. Pancini", University Federico II, Via Cinthia 6, 80126, Napoli, Italy\label{aff37}
\and
Instituto de Astrof\'isica e Ci\^encias do Espa\c{c}o, Universidade do Porto, CAUP, Rua das Estrelas, PT4150-762 Porto, Portugal\label{aff38}
\and
Faculdade de Ci\^encias da Universidade do Porto, Rua do Campo de Alegre, 4150-007 Porto, Portugal\label{aff39}
\and
Dipartimento di Fisica, Universit\`a degli Studi di Torino, Via P. Giuria 1, 10125 Torino, Italy\label{aff40}
\and
INFN-Sezione di Torino, Via P. Giuria 1, 10125 Torino, Italy\label{aff41}
\and
Institute Lorentz, Leiden University, Niels Bohrweg 2, 2333 CA Leiden, The Netherlands\label{aff42}
\and
Leiden Observatory, Leiden University, Einsteinweg 55, 2333 CC Leiden, The Netherlands\label{aff43}
\and
INAF-IASF Milano, Via Alfonso Corti 12, 20133 Milano, Italy\label{aff44}
\and
Centro de Investigaciones Energ\'eticas, Medioambientales y Tecnol\'ogicas (CIEMAT), Avenida Complutense 40, 28040 Madrid, Spain\label{aff45}
\and
Port d'Informaci\'{o} Cient\'{i}fica, Campus UAB, C. Albareda s/n, 08193 Bellaterra (Barcelona), Spain\label{aff46}
\and
Institute for Theoretical Particle Physics and Cosmology (TTK), RWTH Aachen University, 52056 Aachen, Germany\label{aff47}
\and
INAF-Osservatorio Astronomico di Roma, Via Frascati 33, 00078 Monteporzio Catone, Italy\label{aff48}
\and
INFN section of Naples, Via Cinthia 6, 80126, Napoli, Italy\label{aff49}
\and
Institute for Astronomy, University of Hawaii, 2680 Woodlawn Drive, Honolulu, HI 96822, USA\label{aff50}
\and
Dipartimento di Fisica e Astronomia "Augusto Righi" - Alma Mater Studiorum Universit\`a di Bologna, Viale Berti Pichat 6/2, 40127 Bologna, Italy\label{aff51}
\and
Institute for Astronomy, University of Edinburgh, Royal Observatory, Blackford Hill, Edinburgh EH9 3HJ, UK\label{aff52}
\and
European Space Agency/ESRIN, Largo Galileo Galilei 1, 00044 Frascati, Roma, Italy\label{aff53}
\and
Universit\'e Claude Bernard Lyon 1, CNRS/IN2P3, IP2I Lyon, UMR 5822, Villeurbanne, F-69100, France\label{aff54}
\and
Institut de Ci\`{e}ncies del Cosmos (ICCUB), Universitat de Barcelona (IEEC-UB), Mart\'{i} i Franqu\`{e}s 1, 08028 Barcelona, Spain\label{aff55}
\and
Instituci\'o Catalana de Recerca i Estudis Avan\c{c}ats (ICREA), Passeig de Llu\'{\i}s Companys 23, 08010 Barcelona, Spain\label{aff56}
\and
UCB Lyon 1, CNRS/IN2P3, IUF, IP2I Lyon, 4 rue Enrico Fermi, 69622 Villeurbanne, France\label{aff57}
\and
Mullard Space Science Laboratory, University College London, Holmbury St Mary, Dorking, Surrey RH5 6NT, UK\label{aff58}
\and
Departamento de F\'isica, Faculdade de Ci\^encias, Universidade de Lisboa, Edif\'icio C8, Campo Grande, PT1749-016 Lisboa, Portugal\label{aff59}
\and
Instituto de Astrof\'isica e Ci\^encias do Espa\c{c}o, Faculdade de Ci\^encias, Universidade de Lisboa, Campo Grande, 1749-016 Lisboa, Portugal\label{aff60}
\and
Department of Astronomy, University of Geneva, ch. d'Ecogia 16, 1290 Versoix, Switzerland\label{aff61}
\and
INAF-Istituto di Astrofisica e Planetologia Spaziali, via del Fosso del Cavaliere, 100, 00100 Roma, Italy\label{aff62}
\and
INFN-Padova, Via Marzolo 8, 35131 Padova, Italy\label{aff63}
\and
Aix-Marseille Universit\'e, CNRS/IN2P3, CPPM, Marseille, France\label{aff64}
\and
Institut d'Estudis Espacials de Catalunya (IEEC),  Edifici RDIT, Campus UPC, 08860 Castelldefels, Barcelona, Spain\label{aff65}
\and
Institute of Space Sciences (ICE, CSIC), Campus UAB, Carrer de Can Magrans, s/n, 08193 Barcelona, Spain\label{aff66}
\and
School of Physics, HH Wills Physics Laboratory, University of Bristol, Tyndall Avenue, Bristol, BS8 1TL, UK\label{aff67}
\and
NRC Herzberg, 5071 West Saanich Rd, Victoria, BC V9E 2E7, Canada\label{aff68}
\and
Institute of Theoretical Astrophysics, University of Oslo, P.O. Box 1029 Blindern, 0315 Oslo, Norway\label{aff69}
\and
Jet Propulsion Laboratory, California Institute of Technology, 4800 Oak Grove Drive, Pasadena, CA, 91109, USA\label{aff70}
\and
Felix Hormuth Engineering, Goethestr. 17, 69181 Leimen, Germany\label{aff71}
\and
Technical University of Denmark, Elektrovej 327, 2800 Kgs. Lyngby, Denmark\label{aff72}
\and
Cosmic Dawn Center (DAWN), Denmark\label{aff73}
\and
Institut d'Astrophysique de Paris, UMR 7095, CNRS, and Sorbonne Universit\'e, 98 bis boulevard Arago, 75014 Paris, France\label{aff74}
\and
Max-Planck-Institut f\"ur Astronomie, K\"onigstuhl 17, 69117 Heidelberg, Germany\label{aff75}
\and
NASA Goddard Space Flight Center, Greenbelt, MD 20771, USA\label{aff76}
\and
Department of Physics and Astronomy, University College London, Gower Street, London WC1E 6BT, UK\label{aff77}
\and
Department of Physics and Helsinki Institute of Physics, Gustaf H\"allstr\"omin katu 2, 00014 University of Helsinki, Finland\label{aff78}
\and
Universit\'e de Gen\`eve, D\'epartement de Physique Th\'eorique and Centre for Astroparticle Physics, 24 quai Ernest-Ansermet, CH-1211 Gen\`eve 4, Switzerland\label{aff79}
\and
Department of Physics, P.O. Box 64, 00014 University of Helsinki, Finland\label{aff80}
\and
Helsinki Institute of Physics, Gustaf H{\"a}llstr{\"o}min katu 2, University of Helsinki, Helsinki, Finland\label{aff81}
\and
Centre de Calcul de l'IN2P3/CNRS, 21 avenue Pierre de Coubertin 69627 Villeurbanne Cedex, France\label{aff82}
\and
Laboratoire d'etude de l'Univers et des phenomenes eXtremes, Observatoire de Paris, Universit\'e PSL, Sorbonne Universit\'e, CNRS, 92190 Meudon, France\label{aff83}
\and
Aix-Marseille Universit\'e, CNRS, CNES, LAM, Marseille, France\label{aff84}
\and
SKA Observatory, Jodrell Bank, Lower Withington, Macclesfield, Cheshire SK11 9FT, UK\label{aff85}
\and
Dipartimento di Fisica "Aldo Pontremoli", Universit\`a degli Studi di Milano, Via Celoria 16, 20133 Milano, Italy\label{aff86}
\and
INFN-Sezione di Milano, Via Celoria 16, 20133 Milano, Italy\label{aff87}
\and
University of Applied Sciences and Arts of Northwestern Switzerland, School of Engineering, 5210 Windisch, Switzerland\label{aff88}
\and
Universit\"at Bonn, Argelander-Institut f\"ur Astronomie, Auf dem H\"ugel 71, 53121 Bonn, Germany\label{aff89}
\and
INFN-Sezione di Roma, Piazzale Aldo Moro, 2 - c/o Dipartimento di Fisica, Edificio G. Marconi, 00185 Roma, Italy\label{aff90}
\and
Dipartimento di Fisica e Astronomia "Augusto Righi" - Alma Mater Studiorum Universit\`a di Bologna, via Piero Gobetti 93/2, 40129 Bologna, Italy\label{aff91}
\and
Department of Physics, Institute for Computational Cosmology, Durham University, South Road, Durham, DH1 3LE, UK\label{aff92}
\and
Universit\'e C\^{o}te d'Azur, Observatoire de la C\^{o}te d'Azur, CNRS, Laboratoire Lagrange, Bd de l'Observatoire, CS 34229, 06304 Nice cedex 4, France\label{aff93}
\and
Universit\'e Paris Cit\'e, CNRS, Astroparticule et Cosmologie, 75013 Paris, France\label{aff94}
\and
CNRS-UCB International Research Laboratory, Centre Pierre Binetruy, IRL2007, CPB-IN2P3, Berkeley, USA\label{aff95}
\and
Institut d'Astrophysique de Paris, 98bis Boulevard Arago, 75014, Paris, France\label{aff96}
\and
Aurora Technology for European Space Agency (ESA), Camino bajo del Castillo, s/n, Urbanizacion Villafranca del Castillo, Villanueva de la Ca\~nada, 28692 Madrid, Spain\label{aff97}
\and
Institut de F\'{i}sica d'Altes Energies (IFAE), The Barcelona Institute of Science and Technology, Campus UAB, 08193 Bellaterra (Barcelona), Spain\label{aff98}
\and
School of Mathematics, Statistics and Physics, Newcastle University, Herschel Building, Newcastle-upon-Tyne, NE1 7RU, UK\label{aff99}
\and
DARK, Niels Bohr Institute, University of Copenhagen, Jagtvej 155, 2200 Copenhagen, Denmark\label{aff100}
\and
Waterloo Centre for Astrophysics, University of Waterloo, Waterloo, Ontario N2L 3G1, Canada\label{aff101}
\and
Department of Physics and Astronomy, University of Waterloo, Waterloo, Ontario N2L 3G1, Canada\label{aff102}
\and
Perimeter Institute for Theoretical Physics, Waterloo, Ontario N2L 2Y5, Canada\label{aff103}
\and
Institute of Space Science, Str. Atomistilor, nr. 409 M\u{a}gurele, Ilfov, 077125, Romania\label{aff104}
\and
Consejo Superior de Investigaciones Cientificas, Calle Serrano 117, 28006 Madrid, Spain\label{aff105}
\and
Dipartimento di Fisica e Astronomia "G. Galilei", Universit\`a di Padova, Via Marzolo 8, 35131 Padova, Italy\label{aff106}
\and
Caltech/IPAC, 1200 E. California Blvd., Pasadena, CA 91125, USA\label{aff107}
\and
Institut f\"ur Theoretische Physik, University of Heidelberg, Philosophenweg 16, 69120 Heidelberg, Germany\label{aff108}
\and
Institut de Recherche en Astrophysique et Plan\'etologie (IRAP), Universit\'e de Toulouse, CNRS, UPS, CNES, 14 Av. Edouard Belin, 31400 Toulouse, France\label{aff109}
\and
Universit\'e St Joseph; Faculty of Sciences, Beirut, Lebanon\label{aff110}
\and
Departamento de F\'isica, FCFM, Universidad de Chile, Blanco Encalada 2008, Santiago, Chile\label{aff111}
\and
Satlantis, University Science Park, Sede Bld 48940, Leioa-Bilbao, Spain\label{aff112}
\and
Centre for Electronic Imaging, Open University, Walton Hall, Milton Keynes, MK7~6AA, UK\label{aff113}
\and
Infrared Processing and Analysis Center, California Institute of Technology, Pasadena, CA 91125, USA\label{aff114}
\and
Instituto de Astrof\'isica e Ci\^encias do Espa\c{c}o, Faculdade de Ci\^encias, Universidade de Lisboa, Tapada da Ajuda, 1349-018 Lisboa, Portugal\label{aff115}
\and
Cosmic Dawn Center (DAWN)\label{aff116}
\and
Niels Bohr Institute, University of Copenhagen, Jagtvej 128, 2200 Copenhagen, Denmark\label{aff117}
\and
Universidad Polit\'ecnica de Cartagena, Departamento de Electr\'onica y Tecnolog\'ia de Computadoras,  Plaza del Hospital 1, 30202 Cartagena, Spain\label{aff118}
\and
INFN-Bologna, Via Irnerio 46, 40126 Bologna, Italy\label{aff119}
\and
INAF, Istituto di Radioastronomia, Via Piero Gobetti 101, 40129 Bologna, Italy\label{aff120}
\and
Department of Physics, Oxford University, Keble Road, Oxford OX1 3RH, UK\label{aff121}
\and
Instituto de Astrof\'isica de Canarias (IAC); Departamento de Astrof\'isica, Universidad de La Laguna (ULL), 38200, La Laguna, Tenerife, Spain\label{aff122}
\and
Universit\'e PSL, Observatoire de Paris, Sorbonne Universit\'e, CNRS, LERMA, 75014, Paris, France\label{aff123}
\and
Universit\'e Paris-Cit\'e, 5 Rue Thomas Mann, 75013, Paris, France\label{aff124}
\and
ICL, Junia, Universit\'e Catholique de Lille, LITL, 59000 Lille, France\label{aff125}
\and
Department of Physics and Astronomy, University of British Columbia, Vancouver, BC V6T 1Z1, Canada\label{aff126}}

\abstract{The \Euclid Q1 fields were selected for calibration purposes in cosmology and are therefore relatively devoid of nearby galaxies. However, this is precisely what makes them interesting fields in which to search for dwarf galaxies in local density environments. We take advantage of the unprecedented depth, spatial resolution, and field of view of the \Euclid Quick Release (Q1) to build a census of dwarf galaxies in these regions. We have identified dwarf galaxies in a representative sample of 25 contiguous tiles in the Euclid Deep Field North (EDF-N), covering an area of 14.25\,$\deg^2$. The dwarf galaxy candidates were identified using a semi-automatic detection method, based on properties measured by the \Euclid pipeline and released as part of the MER catalogue. A selection cut in surface brightness and magnitude was used to produce an initial dwarf candidate catalogue, followed by a cut in morphology (removing background spirals) and $\IE-\HE$ colour (removing red ellipticals). This catalogue was then visually classified to produce a final sample of dwarf candidates, including their morphology, number of nuclei, globular cluster (GC) richness, and presence of a blue compact centre. We identified 2674 dwarf candidates, corresponding to 188 dwarfs per square degree. The visual classification of the dwarfs reveals a slightly uneven morphological mix of 58\% ellipticals and 42\% irregulars, with very few potentially GC-rich (1.0\%) and nucleated (4.0\%) candidates but a noticeable fraction (6.9\%) of dwarfs with blue compact centres. The distance distribution of 388 (15\%) of the dwarf candidates with spectroscopic redshifts peaks at about 400\,Mpc. Their stellar mass distribution confirms that our selection effectively identifies dwarfs while minimising contamination. The most prominent dwarf overdensities are dominated by dEs, while dIs are more evenly distributed across the field of view. This work highlights \Euclid’s remarkable ability to detect and characterise dwarf galaxies across diverse masses, distances, and environments.}

    
\titlerunning{\Euclid:  Dwarf galaxies in Q1}
\authorrunning{Marleau et al.}
\maketitle

\section{\label{sc:intro} Introduction}

Dwarf galaxies are characterised by their small size and low luminosity, with stellar masses generally smaller than about $10^{9}\,{M}_{\odot}$ \citep{Hodge1971}. This sets the Large Magellanic Cloud, with a stellar mass of $2.7\times10^{9}\,{M}_{\odot}$ \citep{Marel2006}, at the threshold of the dwarf galaxy regime. Dwarf galaxies are the most abundant type of galaxies \citep{Binggeli1988} and in our hierarchical model of structure formation are thought to be the building blocks of the Universe \citep{White1978}. They are distinct from other low-mass objects, such as star clusters, by their larger radii \citep{Kim2016a,Simon2019} and, most importantly, by their dark matter content \citep{Mateo1998,Collins2013,Kim2016b}. This latter key distinction determines whether the object is a dwarf galaxy, which contains dark matter, or instead is a star cluster.

Dwarf galaxies come in several morphological classes. Late-type dwarf galaxies are star-forming galaxies that are split into dwarf irregular (dI or dIrr) and blue compact dwarf (BCD) galaxies, where the main difference between the two is the star-formation rate. Early-type dwarf galaxies are typically quenched, meaning they have ceased star formation. They are classified into dwarf spheroidal (dSph), dwarf elliptical (dE), ultra-diffuse galaxies (UDGs), ultra-compact dwarfs (UCDs), and ultra-faint dwarf (UFD) galaxies, and can be either nucleated or non-nucleated. Some UDGs, especially those in the field, are star-forming late-type objects. The distinction between dwarf ellipticals and dwarf spheroidals is historical in nature, with the early-type dwarf galaxies in our own Local Group dubbed as dSphs \citep[e.g.,][]{Mateo1991,Ibata1994}, while outside, such as in the Virgo cluster, they are usually referred to as dEs \citep[e.g.,][]{Binggeli1994,Lisker2006}. The dSphs are also usually considered to be less massive than the dEs. In this paper the term `dE' refers to all quenched dwarf galaxies.

The morphology of dwarf galaxies appears to be driven by their environments. Early-type dwarf galaxies are found mostly in group and cluster environments \citep{Binggeli1987, Geha2012}, while late-type dwarfs are also found in the field \citep{Papastergis2015,Prole2021,Heesters2023}. In dense environments, various mechanisms can quench a dwarf galaxy \citep{Boselli2008,Kormendy2009,Boselli2022}, such as ram-pressure stripping, tidal harassment, and starvation.  In ram-pressure stripping, the interaction between the hot intracluster gas and the infalling dwarf galaxy removes its star-forming gas reservoir \citep{Gunn1972}. In tidal harassment (or tidal stripping) the gas is removed via tidal deformations, usually forming two arms \citep{Moore1996,Mayer2001,Smith2010}. Starvation describes the mechanism where a galaxy is depleted of its star-forming gas in a cluster environment \citep{Larson1980}. Although these mechanisms are dominant in cluster environments, a recent survey of dwarf galaxies has identified dEs in large numbers in the low- to medium-density environments of groups of galaxies \citep[e.g.,][]{Habas2020}, which suggests that there is likely more than one pathway to their formation.

The detection of faint dwarf galaxies is no trivial task, due to their low surface brightness and relatively small sizes. In the past, such surveys have been heavily based on visual inspection of astronomical images \citep[e.g.,][]{Karachentsev2015,Ordenes-Briceno2016,Wittmann2017,Park2017,Mueller2017,Paudel2023,Crosby2024,Montes2024}, limiting the area of the night sky that can effectively be covered by such a census. Semi-automated approaches, using detection tools such as \texttt{SourceExtractor} \citep{Bertin1996}, \texttt{NoiseChisel} \citep{Akhlaghi2015}, or \texttt{MTObjects} \citep{Teeninga2015}, have yielded catalogues of potential dwarfs \citep[e.g.,][]{Merritt2016,Venhola2017,Carlsten2020,Habas2020,Tanoglidis2021}, which, however, still need to be inspected by eye to sort out false positives, and which often miss out obvious dwarf galaxies near bright sources such as stars or galaxies \citep{Mueller2020b}. Currently, no fully automated process exists that yields both a pure and complete dwarf galaxy catalogue without visual inspection. This is a problem for wide-field surveys such as \Euclid, which cover large fractions of the entire night sky.

The \Euclid space mission \citep{Laureijs11,EuclidSkyOverview} is set to observe nearly one-third of the sky across four photometric bands, extending from the red optical to the near infrared (NIR). The telescope is mounted with two key instruments: the visible instrument (VIS; \citealt{EuclidSkyVIS}), with one single broad-band filter, $\IE$, devoted to imaging at red optical wavelengths; and the Near Infrared Spectrometer and Photometer (NISP; \citealt{Schirmer-EP18,EuclidSkyNISP}), which provides both imaging in the NIR through three filters ($\YE$, $\JE$, and $\HE$), and low-resolution NIR spectroscopy. In addition to providing images with superb spatial resolution (VIS: \ang{;;0.20}, NISP: $\simeq \ang{;;0.48}$; see Sect.\ref{sc:eucliddata}) and covering a large part of the sky (14\,000\,deg$^{2}$; \citealt{Scaramella-EP1}), \Euclid is poised to reach unprecedented surface brightness sensitivity in the different surveys ($29.8$\,mag\,arcsec$^{-2}$ for the Euclid Wide Survey (EWS; \citealt{Scaramella-EP1} and 31.8\,mag\,arcsec$^{-2}$ for the Deep Survey).

Thanks to all this, \Euclid is expected to detect and characterise of the order of a million new dwarf galaxies, including tens of thousands of new UCDs and UDGs, over a range of distances and environments. UDGs are particularly interesting, given the ongoing debate on their nature, formation mechanisms, and variety of globular cluster (GC) populations compared to the general population of dwarf galaxies (\citealp{Lim2018,Forbes2020,Marleau2021,Gannon2022,Saifollahi2022,Ferre-Mateu2023,Marleau2024}; Saifollahi et al.\ 2025, in prep.). 

This new era of wide-field imaging surveys, such as the EWS, is set to be transformational in the way we understand galaxy formation and evolution, particularly at the low-mass end of the galaxy mass function. Already with the large number of Early Release Observations (ERO; \citealt{EROcite}) papers, \Euclid has proven its outstanding performance on all the scientific topics involved. But in particular, it has shown its outstanding capability to detect new dwarfs and reveal their GC systems (\citealp{EROPerseusDGs, EROFornaxGCs, EROPerseusOverview}; Saifollahi et al.\ 2025, in prep.). Examples of these exquisite capabilities are shown in the ERO \citep{EROcite} papers describing the observation and characterisation of the dwarf galaxies in the Perseus galaxy cluster (\citealp{EROPerseusDGs}; Saifollahi et al.\ 2025, in prep.), the GCs in the Fornax cluster \citep{EROFornaxGCs,EP-Voggel}, the diffuse stellar structures and GCs in the Dorado group (Urbano et al.\ 2025, in prep.), and the ERO project on nearby galaxies \citep{ERONearbyGals}. 

In this paper, we present the detection and characterisation of the dwarf galaxy population in a representative sample of the \cite{Q1cite}. The Q1 data are outlined in Sect.~\ref{sc:data}. Sections~\ref{sc:bkgsub}, \ref{sc:colour} and \ref{sc:cat} detail the data preparation procedure and the catalogue used for source detection and photometric analysis. Section~\ref{sc:method} describes the methodology for processing the data and identifying the dwarf galaxy candidates. Section~\ref{sc:results} presents the results of the visual classification, photometric and structural properties, spectroscopic redshift, and spatial distribution of the dwarf candidates. Finally, Sect.~\ref{sc:conclusion} summarises the findings on the dwarf population in the Q1 tiles.

\section{\label{sc:data} Data} 

\subsection{\label{sc:eucliddata} Euclid observations} 

The Q1 data span $63.1\,\deg^2$ \citep{Q1-TP001} and cover three Euclid Deep Fields (EDFs; $53\,\deg^2$), with a single visit depth equivalent to one Reference Observation Sequence (ROS), similar to the Euclid Wide Survey (EWS, \citealp{Scaramella-EP1}). These three fields are knows as EDF North (EDF-N), EDF South (EDF-S), and EDF Fornax (EDF-F). A detailed description of the EDFs can be found in \citet{EuclidSkyOverview}. The Q1 data also include observations of the Lynds Dark Nebula LDN1641; the seven ROSs of this field were obtained as part of the commissioning of the Fine Guidance Sensor (FGS).

The FoV of each Q1 tile is of size 0.28\,deg$^{2}$ ($\ang{;32;}\,\times\,\ang{;32;}$; \citealt{Q1-TP004}). The data were obtained in a dithered observation sequence where an image in \IE\ is taken simultaneously with slit-less grism spectra in the NIR, followed by NIR images taken through \JE, \HE, and then \YE. The telescope is then dithered and the sequence is repeated four times. This observation sequence is similar to the ROS that is being used to observe the EWS. The integration time per dither position is 549.6\,s in the \IE\ filter and 87.2\,s in each of the \YE, \JE, and \HE\ filters \citep{EuclidSkyOverview}. 

\begin{figure*}[ht!]
\centerline{
\includegraphics[width=\linewidth]{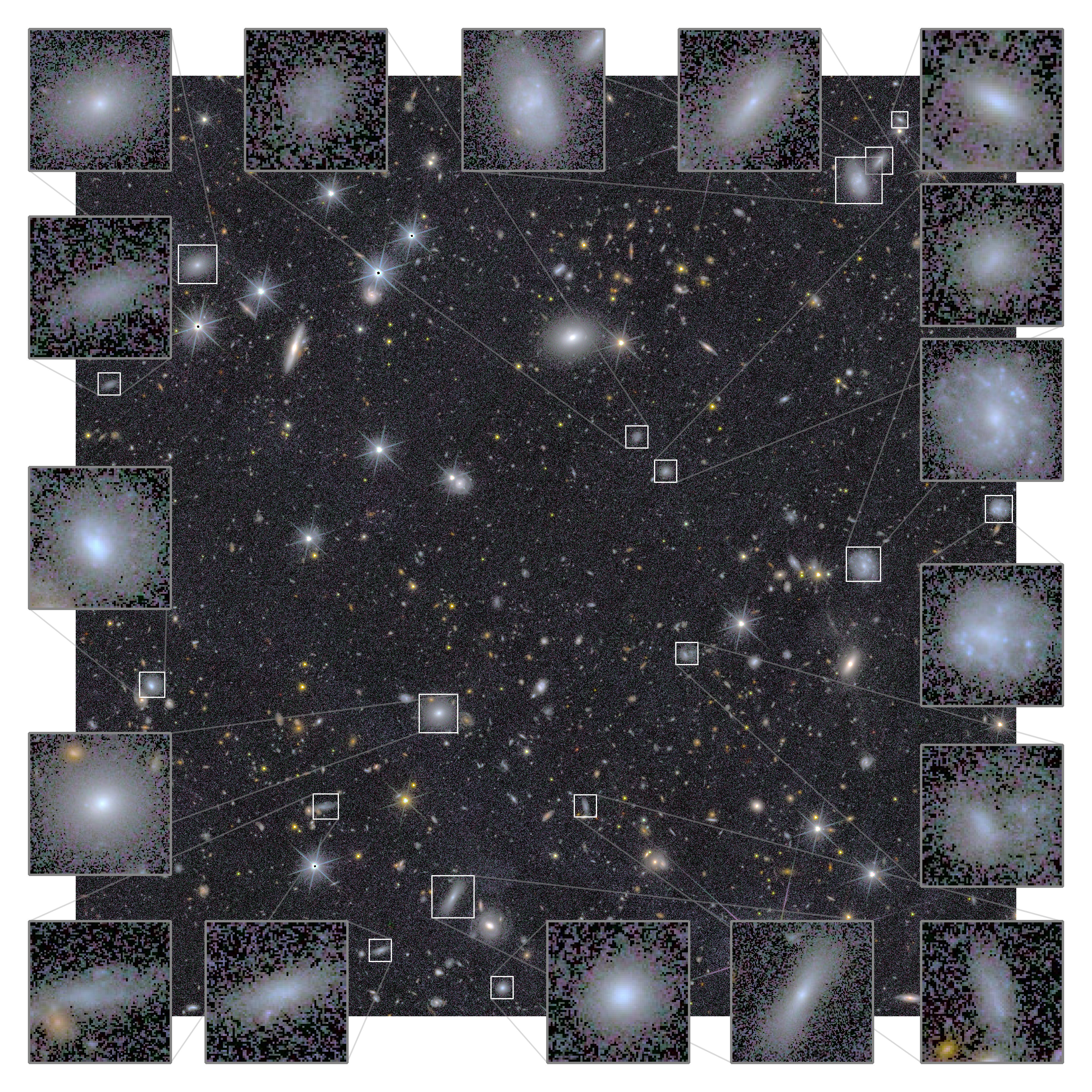}
}
\caption{VIS-NISP colour image of a zoomed-in region of the EDF-N tile 102158889, \ang{;5.2;} $\times$ \ang{;5.2;} in size, created using the \IE\ band in blue, the \YE\ band in green, and the \HE\ band in red. The colour image has the same spatial resolution as the high-resolution \IE\ band image. Examples of seventeen dwarf galaxy candidates with score $\geq 0.67$ and area $\geq 700$\,pixel$^2$ (see the text for details) are shown in the individual cutouts of sizes between \ang{;;5} $\times$ \ang{;;5} and \ang{;;15} $\times$ \ang{;;15}. In all images, north is up and east is to the left.}
\label{fig:tilecol}
\end{figure*}

\begin{figure*}[ht!]
\centerline{
\includegraphics[width=0.9\linewidth]{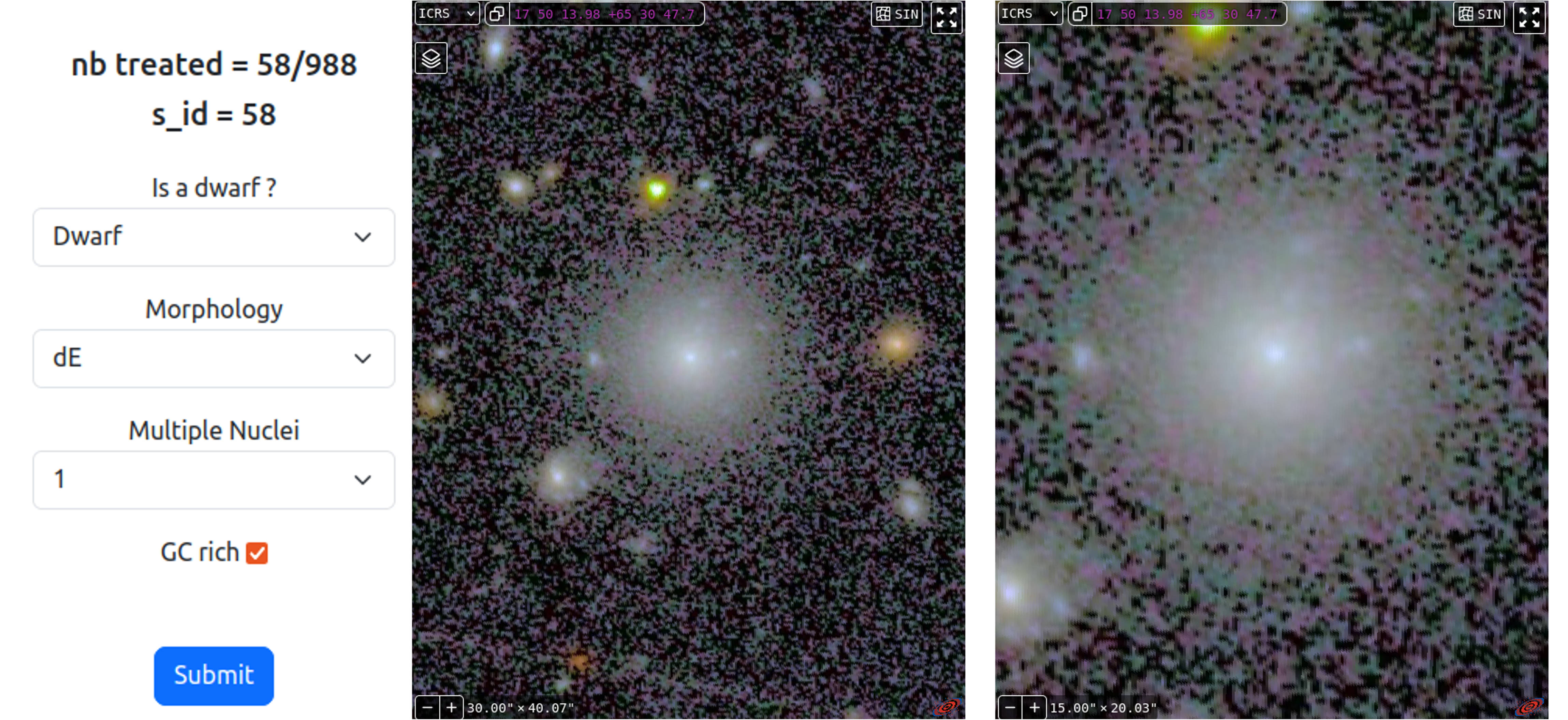}}
\caption{The online tool (Jafar) used to validate and visually classify the dwarf galaxy candidates, showing an example of a nucleated and potentially GC-rich dwarf galaxy candidate.}
\label{fig:classification}
\end{figure*} 

\begin{figure}[ht!]
\centerline{\includegraphics[width=\linewidth]{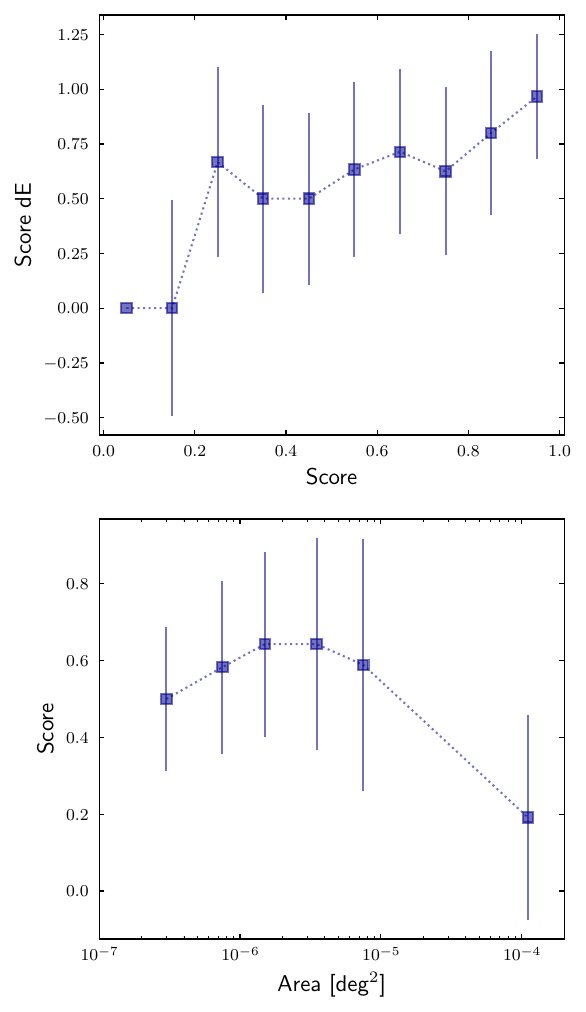}}
\caption{\textit{Top:} Score that the dwarf is classified as a dE as a function of the score that the candidate is a real dwarf. \textit{Bottom:} Median score as a function of the source area from the annotation tool. The error bars represent the standard deviation in each bin. This plot provides the range of object sizes that we could more securely consider for the dwarf validation. It shows that early-type dwarfs get a better consensus and that too small or large objects tend to be excluded as dwarfs.}
\label{fig:scores}
\end{figure}

The pixel sizes for the VIS and NISP instruments are \ang{;;0.1} and \ang{;;0.3}, respectively, which means that for both instruments the point-spread function (PSF) is slightly undersampled. In the final Q1 stacked frames (OU-MER), VIS and NISP images are re-sampled to a similar pixel size of \ang{;;0.1}. The final Q1 data have a median PSF full width at half maximum (FWHM) of \ang{;;0.20}, \ang{;;0.48}, \ang{;;0.50}, and \ang{;;0.54} in \IE, \YE, \JE, and \HE, respectively \citep{Q1-TP004}. The zero point (ZP) in all filters is ${\rm ZP} = 23.9\,{\rm AB\,mag}$. The details of the data reduction are described in \citet{Q1-TP002} and \citet{Q1-TP003}. Hereafter, we refer to AB magnitudes as simply magnitudes.

In total, the Q1 data release contains 351 tiles. However, the seven tiles pertaining to LDN1641 display a large amount of foreground Galactic extinction which leads to fewer sources being detected and a larger extinction correction for the photometric measurements. Therefore, we excluded these seven tiles from our analysis, leaving 344 Q1 tiles. Overall, a total of 57 tiles have bad quality (e.g., extreme edge, no colour information, and dark nebula) and 91 tiles have partial coverage. This leaves a number of 351$-$57$-$91$=$203 tiles with good quality data and full coverage. 

In this paper, we present the detection and characterisation of the dwarf galaxies in a representative region of the Q1 data release, consisting of a contiguous set of 25 Q1 tiles in the EDF-N. One of the reason that led us to select the tiles in the EDF-N field is that the EDF-N has very good spectroscopic redshift coverage. The tiles that we selected have labels starting with 10215, followed by: 9778, 9777, 9776, 9775, 9774, 9490, 9489, 9488, 9487, 9486, 9194, 9193, 9192, 9191, 9190, 8893, 8892, 8891, 8890, 8889, 8587, 8586, 8585, 8584, and 8583. The analysis of the remainder of the Q1 tiles will be presented in a separate future paper. 

\subsection{\label{sc:bkgsub} Background restoration} 

The \Euclid standard pipeline includes background-removal steps that are optimised for its cosmological science, which requires the shape measurement for weak lensing analysis of millions of small (1\arcsec--2\arcsec) distant galaxies. Structures larger than the mesh size used for the background determination are erased by the standard pipeline. The larger the source, the higher the effect of the background subtraction. 

For the background-subtracted image by the MER processing pipeline (BGSUB MER product; \citealt{Q1-TP004}), even at scales of a few 10\arcsec, i.e., the scale of dwarf galaxies, the local background subtraction is detrimental. Therefore, the final MER background subtraction step was reversed by adding back the background measured by the MER processing pipeline (BGMOD MER product) to the BGSUB MER product, hence providing the image before the final MER background removal. As a result, these MER-background restored images are stacks of VIS images that have had their background subtracted by the VIS processing pipeline, which provides good background subtraction up to about 1\arcmin\ scale. This is not the case for the Q1 NIR images, which have a background subtraction that has not been optimised for analysis of low surface brightness features at this scale. The rendering of the colour images shown in this paper is affected by this for the most extended objects.

It is important to note, however, that the MER catalogue \citep{Q1-TP004}, discussed in more detail in Sect\,\ref{sc:cat}, was generated from the standard pipeline produced background subtracted image. This means that the photometric parameters (e.g., magnitudes, surface brightness) will likely be impacted. However, for the small dwarfs, we do not expect this impact to be significant (the effect becomes more important for the bright extended galaxies).

\subsection{\label{sc:colour} Colour images} 

Colour images at the VIS spatial resolution were generated for all 344 tiles using the AstrOmatic sofware program {\tt STIFF}, a program that converts scientific FITS images to the more popular TIFF format for illustration purposes \citep{Bertin_2012}. This was done using the \IE, \YE, and \HE images, since the \JE image is the most affected by persistence effects because it is taken right after the NISP spectroscopic data. An example colour image of a zoomed-in region of tile 102158889 is shown in Fig.\,\ref{fig:tilecol}. This image illustrates that dwarf galaxy candidates are easily identifiable in the Q1 data set.

The VIS-NISP colour images were then generated in the {\tt HiPS} format \citep{Fernique2015}, for visualisation and classification of the dwarf galaxy candidates, using the \IE band in blue, the \YE band in green, and the \HE band in red. The colour image has the same spatial resolution as the high-resolution \IE\ band image. To produce these images, the luminance channel was taken from the \IE\ band image, while the \IE, \YE, and \HE\ images were used for the three colour channels (corresponding to blue, green, and red, respectively). As the NIR images are simply resampled at the VIS resolution, this means that the colour images are limited by the NIR resolution.

\subsection{\label{sc:cat} Source catalogue}

The source catalogue was taken directly from the MER catalogue. This product is an output from the MER Catalogue Assembly Processing Element \citep{Q1-TP004}. The MER catalogue is the final merged catalogue, with photometric and morphological information. It contains object positions, total fluxes, colours and some object characterisation parameters. The catalogue information for each tile is stored in a FITS file.

For all sources in the MER catalogues, the magnitudes of the objects were derived by taking the fluxes reported in the MER catalogues and applying the zeropoint given above. The extinction correction was applied using the following formula: $m_0 = m - R_{\lambda} \times \rm{GAL\_EBV}$, where 'GAL\_EBV' is the estimated Galactic $E(B-V)$ at the source centroid according to the reference {\it Planck} map provided in the MER catalogue and $R_{\lambda}$ is taken from the Milky Way extinction law ($R_V = 3.1$; \citealt{Fitzpatrick1999}).

Comparison of the MER catalogue with our own run of \texttt{SourceExtractor} on one of the tiles showed the MER catalogue to be very complete and hence we decided to rely on the MER catalogue for our analysis. Using a methodology for detecting dwarf galaxies based on the source catalogue generated by the pipeline has two main advantages: firstly, it makes the analysis more easily reproduceable since the catalogues are publicly available; and secondly, it avoids unnecessary computing time, which will become even more important when we are dealing with the large number of tiles in the \Euclid Data Release 1 (DR1). It is important to note that the MER catalogue associated with each tile does not contain sources from the edge region (\ang{;;60} in size) in order to avoid the duplication of sources between adjacent tiles in the overlap regions.

We also investigated the completeness of the MER detection to determine whether the most nearby and extended dwarf galaxies are adequately detected and included in the final source catalogue. Given the lack of such nearby dwarf galaxies within 50\,Mpc in the EDF-N, we focused our assessment on the dwarf galaxies found in EDF-S. We searched the SIMBAD database (\citealp{simbad}) for such nearby objects and found six dwarf galaxies within the EDF-S with radial velocities between 700 and 3500\,\kms, corresponding to redshift distances of 10 to 50\,Mpc, respectively. All six of these objects have been detected and included in the MER catalogue.

\section{\label{sc:method} Dwarf identification and classification}

To efficiently identify dwarf candidates in the Q1 tiles, we adopted a semi-automated methodology. The procedure we followed consisted firstly of selecting a representative tile, namely a `reference' tile, for which we performed a full visual detection and classification analysis. The results obtained for the reference tile were then used to calibrate our selection method for all Q1 tiles. 

The general steps of our methodology were divided into two main parts. First, our analysis of the reference tile consisted of: (i) visually inspecting the colour image and annotating the dwarf candidates; (ii) comparing and merging the catalogues of the different classifiers; (iii) visually classifying objects in the merged catalogue (16 classifiers per object); and (iv) creating the final catalogue from the average score selection cut. This reference sample of dwarf galaxies was then used to define the selection cut for all of the Q1 tiles. The second part of our analysis therefore consisted of: (i) applying the selection cut to all of the tiles and automatically generating the candidate list; (ii) visually classifying the dwarf candidates (three classifiers per object); (iii) comparing and merging the catalogues of the different classifiers; (iv) creating the catalogue of the dwarfs in all the tiles from the average score selection cut; and (v) visually inspecting the tiles and catalogued dwarfs and creating the final catalogue. These two main parts of our analysis are described in more detail in the following sections.

\begin{figure*}[ht!]
\centerline{
\includegraphics[width=\linewidth]{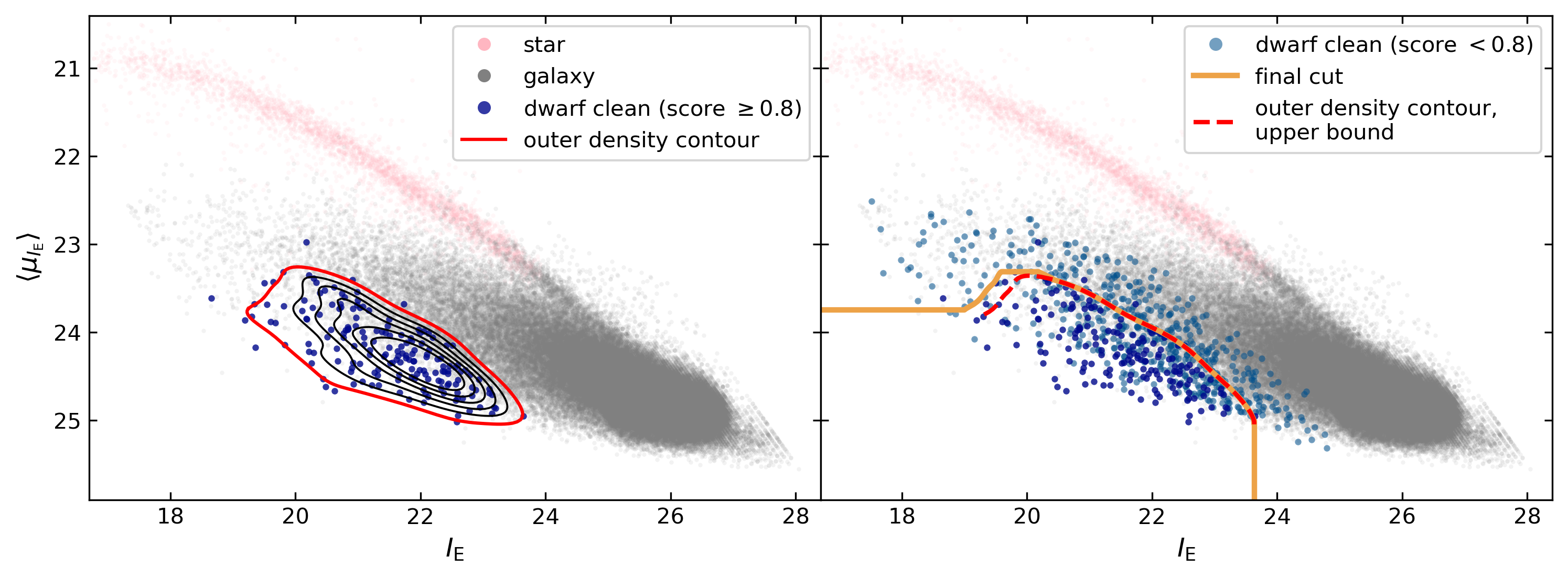}}
\caption{\textit{Left:} Isodensity contour plot in average surface brightness versus apparent magnitude space for the visually identified dwarf galaxy candidates in the reference tile 102158889 with a score $\geq 0.8$. The dwarf candidate sample was manually cleaned to ensure good MER photometry of the sources. The upper portion of the outermost contour is used as basis for the subsequent automated selection cut (\textit{right}). The non-dwarf galaxies (\texttt{POINT\_LIKE\_PROB} $< 0.8$"; grey points) and stars (\texttt{POINT\_LIKE\_PROB} $\geq 0.8$"; pink points) are also included for reference. \textit{Right:} An example of the automatic selection cut in the reference tile. We extended slightly the upper portion of the outer density contour (red) to select the few bright dwarf candidates with score $\geq 0.8$ that would otherwise have been missed. The final selection cut is shown in orange. Many of the less confident dwarfs (score $< 0.8$) are also selected by this line.}
\label{fig:selection}
\end{figure*}

\begin{figure}[ht!]
\centerline{
\includegraphics[width=\linewidth]{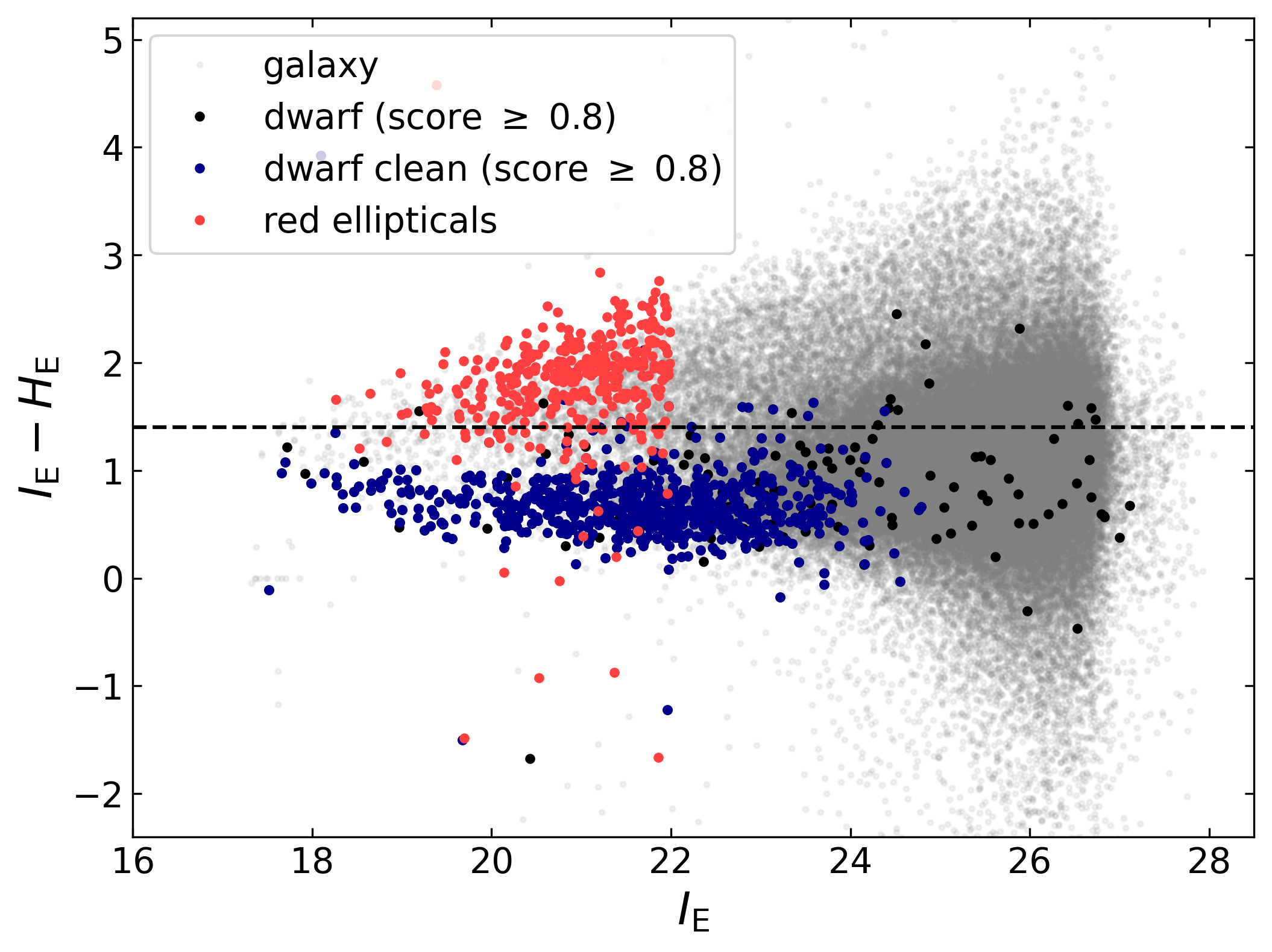}}
\caption{Colour-magnitude diagram showing the colour cut of $I_{\rm E} - H_{\rm E} \leq 1.4$ used to remove the red elliptical galaxies from our automatically selected dwarf sample. The MER detections in the reference tile are in grey and the objects in the visual catalogue for score $\geq 0.8$ are in black. The manually cleaned visual dwarfs for score $\geq 0.8$, as described in Fig.\,\ref{fig:selection}, are in blue-violet, and the visually identified red elliptical galaxies are in red.}
\label{fig:colorcut}
\end{figure}

\subsection{Reference tile: visual identification}

To ensure the robustness of our selection criteria, we first created a visual catalogue of dwarf galaxies that could be used as a reference sample.

We used the \texttt{Jafar} annotation tool \citep{Sola2024} to identify and validate the dwarf candidates in tile 102158889, carefully selected as representative of the Q1 data set. \texttt{Jafar} allows one to navigate across images, zoom in and out, and precisely delineate the contours of any objects of interest, while also attaching a corresponding label. The coordinates, labels, and geometrical properties of the annotated shapes are stored in a centralised database. This identification and annotation process is similar to the one used in \citet{EROPerseusDGs} to determine the catalogue of dwarf galaxy candidates in the ERO data for the Perseus galaxy cluster, and is divided into several steps. 

First, 16 classifiers were presented with an inpainted coloured version of the tile 102158889 with the \IE pixel scale, as described in Sect.~\ref{sc:colour}. The combined information from colour and high spatial resolution enabled the distinction between background objects, artefacts, and dwarf galaxy candidates. Each classifier independently inspected the full tile, annotating potential dwarf galaxies. These candidates were delineated using ellipses, while internal substructures such as nuclei were annotated using circles. 

The sixteen individual catalogues contained a total of 2759 annotations. These classifications were gathered into a single table by removing duplicate objects; each catalogue was first cross-matched against itself in order to identify and remove mislabelled structures (e.g., nuclei that were erroneously flagged as galaxies), then we identified duplicates within a search radius that was scaled by the square root of the annotated area of each galaxy. This allows for larger tolerances around more extended galaxies, yet should ensure that small neighbouring dwarf candidates are not falsely merged into a single object. This generated a list of 989 objects of interest that had been annotated by at least one individual, 463 had been flagged by at least two people, and 146 galaxies had been flagged by a majority of the classifiers at this early stage. 

For galaxies that were flagged more than once, we adopt the average RA and Dec as the final coordinates of the object. To test how accurate this value is, we performed a final cross-match between the catalogue containing the objects of interest and the original annotations, using the square root of the averaged area as the new search radius. The separation between the averaged coordinates and the original annotated coordinates peaks at \ang{;;0.2} and drops to almost zero at about \ang{;;1.5}, with a nearly negligible tail extending to larger radii. In general, we find very good agreement between the annotations.

\subsection{\label{sc:visclass} Reference tile: visual classification}

The next step was to validate the annotated objects, i.e., to visually re-inspect all objects of interest to confirm a sample of dwarf candidates and reject any interlopers among the 989 flagged galaxies. This task was also performed in \texttt{Jafar}, but with a different interface. Instead of navigating around the full image, the 16 classifiers were now presented with two panels: one displaying a \IE\YE\HE coloured image centred on the object with a fixed field-of-view of $30\farcs \times 24\farcs 5$, and the other displaying a zoomed-in version of the image (see Fig.\,\ref{fig:classification}). These displays show a subset of the full image, which could still be manipulated in all the ways the images could be adjusted during the earlier annotation stage, but also allowed classifiers to display data from other surveys (e.g., Pan-STARRS DR1) if more information was needed. However, the initial displays were chosen to facilitate quick decisions for nearly all of the candidates. The classifiers were asked if the object was considered a dwarf galaxy and to assess the morphology (`dE', early-type galaxy; `dI', star-forming irregular galaxy; `disturbed', tidally disrupted or interacting galaxy), GC-richness, and number of nuclei. It is important to note that the galaxies classified as disturbed were explicitly dEs; hence, there is no disturbed dI.

\begin{figure*}[ht!]
\centerline{
\includegraphics[width=\linewidth]{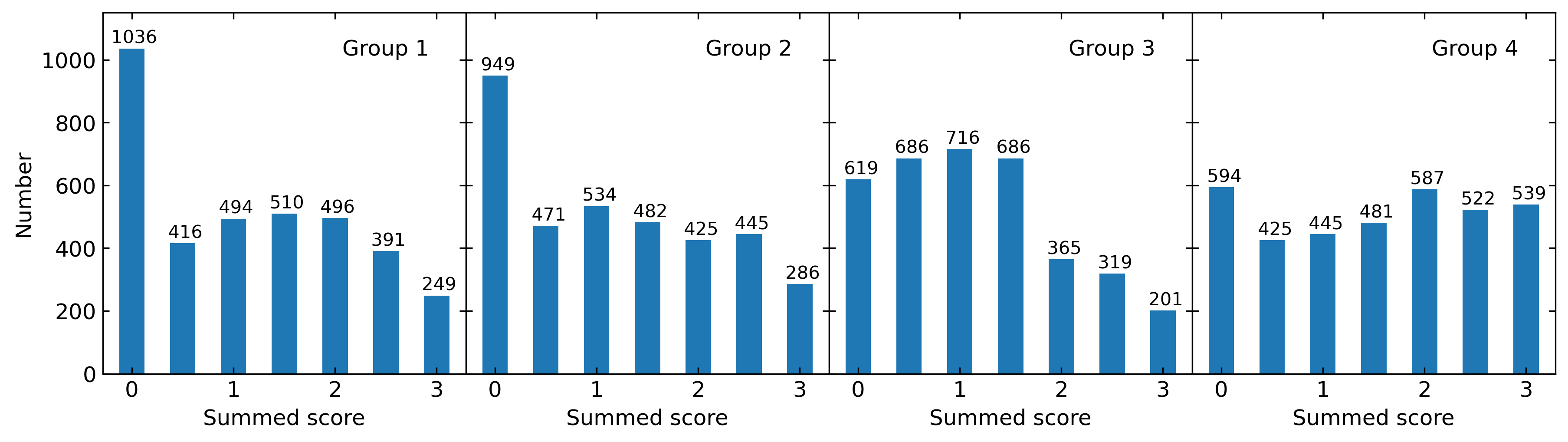}}
\caption{Summed scores (`dwarf' $=1$, `unsure dwarf' $=0.5$, `not dwarf' $=0$) for the three classifiers in each of the four groups of classifiers. The groups were arranged to avoid combinations of classifiers who were consistently more conservative (or more liberal) in the reference tile. Some differences still remain between the groups, but these are not severe for the more confident dwarf candidates, and this may reflect a degree of cosmic variance between regions on the sky.  }
\label{fig:group_classifications}
\end{figure*}

\begin{figure}[ht!]
\centerline{
\includegraphics[width=0.95\linewidth]{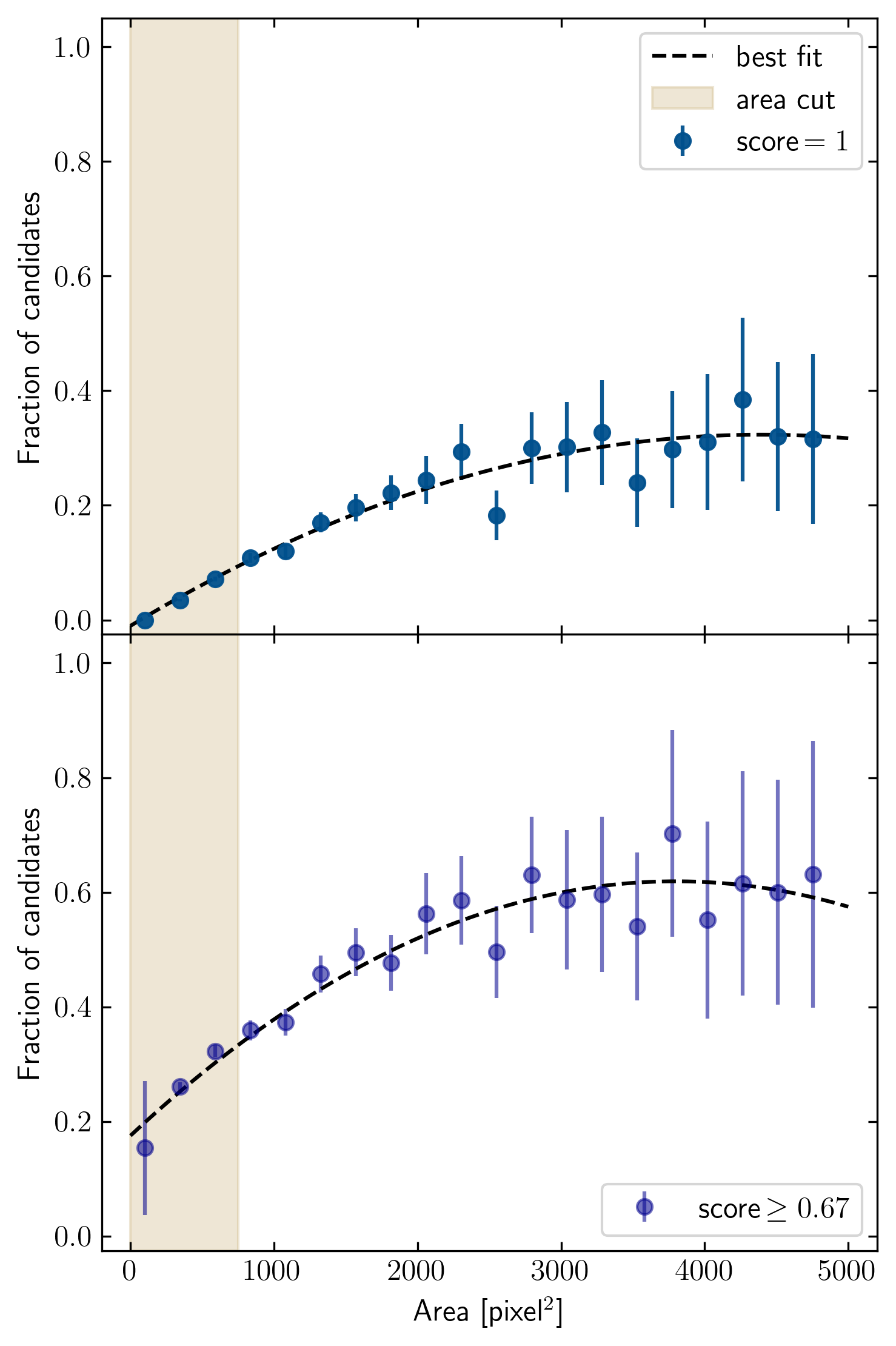}}
\caption{\textit{Top:} Fraction of all dwarf candidates, from all groups, with a score = 1 per area bin. A quadratic function was fit to the data to create the best-fit line. For this paper, we have dropped candidates with areas $<700$\,pixel$^2$, as highlighted by the tan region. The errors were propagated assuming Poisson statistics and that the number of dwarf candidates and the total number of galaxies per bin are independent. \textit{Bottom:} Same as top, but for all dwarfs in the final sample (score $\geq 0.67$).}
\label{fig:sizecut}
\end{figure}

A set of guidelines was given to the users for consistency: (i) if the object looks like a dwarf elliptical but has a prominent red nucleus, this is likely a background elliptical; (ii) if the object looks like a dwarf elliptical but has a prominent blue central clump, this is likely a real dwarf (transitional quenched object); (iii) if the object shows spiral structures, this is likely a background disc; (iv) if the object has blue clumps, but a redder nucleus and underlying reddish diffuse stellar disc, this is likely a background massive clumpy disc; (v) if the object has the same apparent size as very nearby obvious background galaxies, this is likely a member of this background group and not a dwarf.

When visually classifying dwarf candidates, one open question was the nature of diffuse galaxies with extended bulges. It can be unclear if these are dwarf galaxies with atypically bright centres, background S0s, or another class of objects. To ensure we did not bias our sample by erroneously including or excluding these galaxies, we further inspected these objects, as described below.

\begin{figure*}[ht!]
\centerline{
\includegraphics[width=\linewidth]{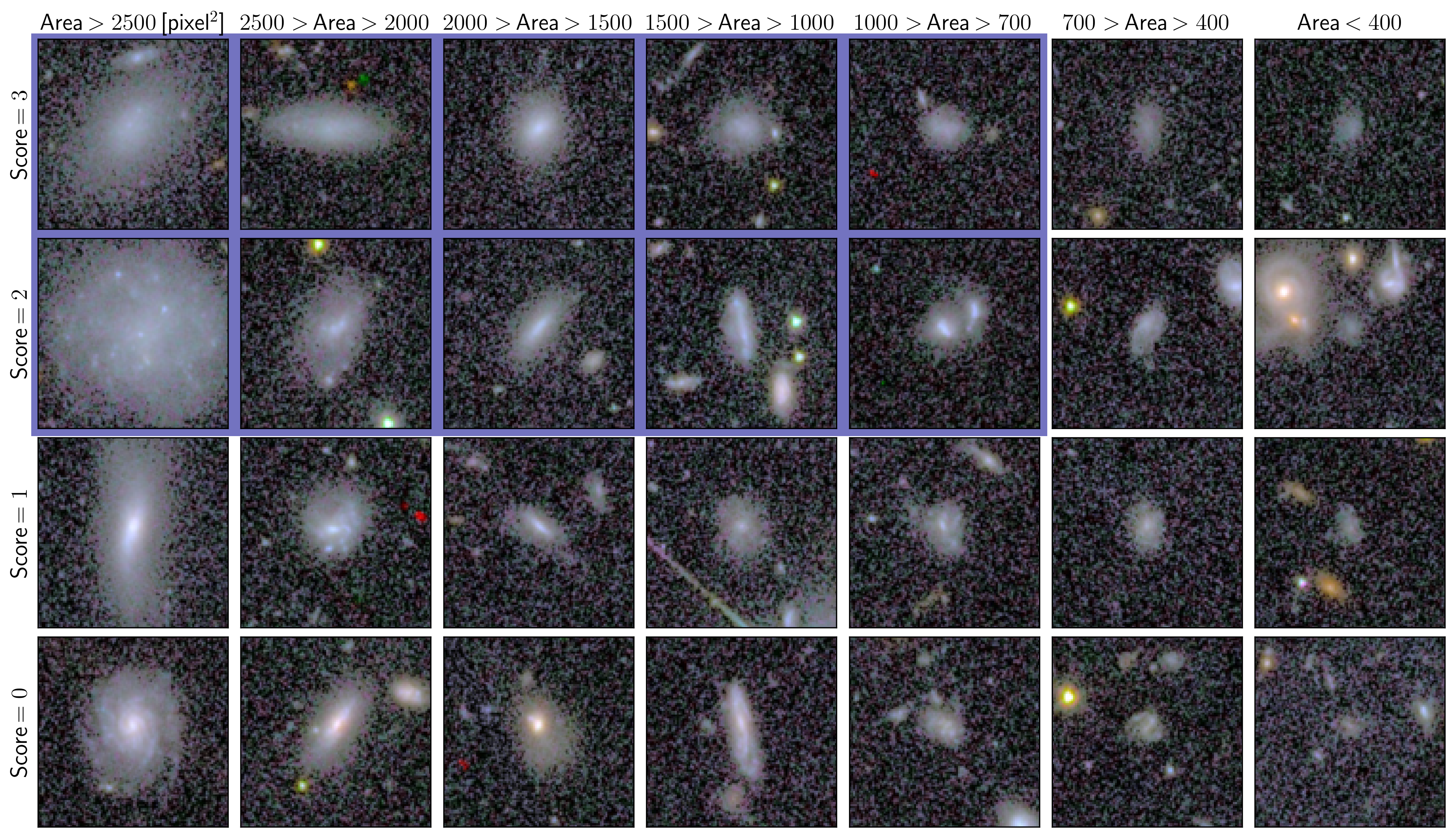}}
\caption{Examples of galaxies returned by the selection cut, as a function of area and visual classification score. The rows and columns with blue-violet background are part of our final sample. The two right-most columns were dropped due to their small sizes and the bottom two rows were dropped for their low classification scores; see Sect.\,\ref{sc:allQ1tilesclass} for details. To better sample the range of areas in the sample, note that the area bins do not have a fixed width. The images are approximately \ang{;;15} $\times$ \ang{;;15} in size, with north up and east to the left.}  
\label{fig:dropouts}
\end{figure*}

In the reference tile, we detected 81 (8.2\%) dwarf galaxies with morphology similar to dEs but with a bright central blue or white core. Similar objects have been identified in recent works as `blue-cored' dwarf elliptical galaxies (e.g., \citealt{Lisker2006,Urich2017,Paudel2020,Jiwon2023}). However, the lack of spatial resolution from ground-based data means that many of the central blue clumps appear as unresolved cores in these images. Early-type galaxies with a large bright core component in the inner regions were introduced by \citet{Sandage1984} and \citet{Binggeli1991} as dwarf lenticular galaxies (dS0s). The surface brightness profile of these dwarfs can be well fit with exponential or King profiles, with an excess of light in the central region \citep{Binggeli1991, Binggeli1993}. However, these galaxies exhibit a variety of disc morphologies, and there is an ongoing debate regarding whether they can be considered a subclass of dEs, due to their similarity in surface brightness, or if they represent a distinct class of dwarf galaxy \citep{Ryden1999}. In the Q1 images, the early-type dwarfs with large cores clearly exhibit a recent or on-going star formation in these regions, and they may be considered transitional dwarf galaxies, an intermediate product of the transformation from gas-rich late-type dwarf galaxies into gas-poor early-type dwarfs. For the rest of the paper we therefore refer to these dwarf galaxies as `transitional dwarf galaxies'.

In order to determine if the classifiers should identify the transitional galaxies as dwarf candidates, we examined those found in the reference tile with spectroscopic redshifts from which we could compute their absolute magnitudes. In the reference tile, we cross-matched the 81 transitional dwarf galaxies with the spectroscopic redshift catalogue of DESI spectra \citep{Desi2024} and found 30 (37$\%$) matches. Aperture photometry was performed for all these candidates using the Python package \texttt{Photutils} \citep{Bradley2024} and good agreement was found with the MER pipeline magnitudes, in particular with the values determined from the \texttt{FLUX$\_$DETECTION$\_$TOTAL} parameter. The early-type dwarfs with bright cores and spectroscopic redshifts were found to have absolute magnitudes in the range $-19 < M(\IE) < -16$. Therefore, we concluded that they all qualify as dwarfs. 

A score was computed for each source with the following weights: $1=$ dwarf is validated; $0 = $ dwarf is not validated; and $0.5 = $ unsure. The total score for each object is the sum of all votes, normalised by the number of classifiers. We obtained similar scores for the morphology, nucleated status, and GC-richness, although these did not have an option for `unsure'. To test our confidence at classifying various types of galaxies, we show the total score as a function of the probability that the dwarf is classified as a dE in Fig.\,\ref{fig:scores} (top panel), and as a function of the annotated area in Fig.\,\ref{fig:scores} (bottom panel). The error bars represent the standard deviation in each bin. From these plots, it can be seen that we are more confident in classifying dE-type dwarfs, while the smallest and largest galaxies were more likely to be rejected.

In \citet{EROPerseusDGs}, a score threshold could be clearly selected to define the final sample of dwarf candidates, due to the uniformity in terms of morphology and distance of the dwarf population in the ERO Perseus images. The selection cut was not as straightforward for the dwarf classification in the Q1 reference tile (and all Q1 tiles), since the dwarf population consists of galaxies located at a range of distances and is made up of a higher fraction of dIs. As our final robust dwarf candidate galaxy sample, we therefore chose a total validation score $\geq 0.8$ to conservatively select our most robust dwarf galaxy candidates. 

\subsection{Reference tile: comparison of visual and MER catalogue}

As discussed earlier, the MER catalogue uses background-subtracted images, so there was some question if it would be appropriate for identifying LSB dwarf candidates. We compared the visual catalogue with the MER catalogue of the reference tile and found a good agreement in the galaxies detected. Using the area measured from the visual annotation of users and the MER catalogue parameter \texttt{SEGMENTATION\_AREA}, we also compared their measured sizes and also found, in general, a good agreement. Moreover, we identified only few situations where the MER catalogue was inconsistent with the visual identification. These were either when the dwarf galaxy candidate was a dI with star-forming clumps and these clumps had been detected as separate sources in the MER catalogue, or when the dwarf galaxy candidate was near a bright galaxy/star and missing from the MER catalogue. 

\begin{figure*}[ht!]
\centerline{\includegraphics[width=0.813\linewidth]{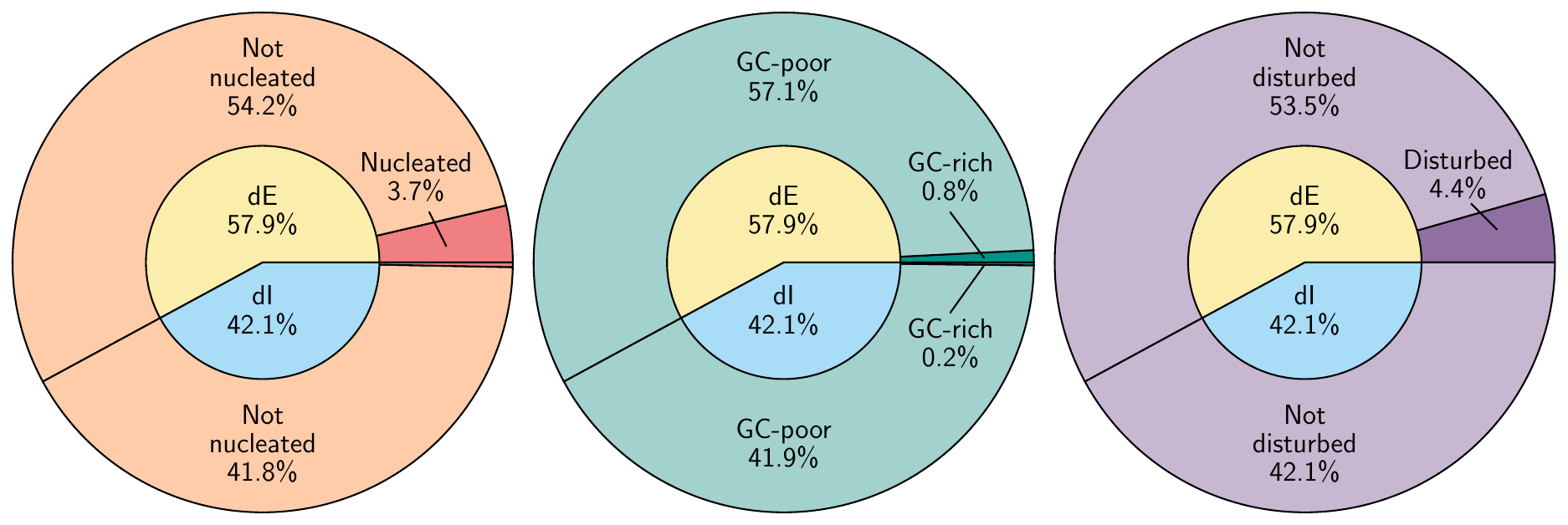}}
\centerline{\includegraphics[width=0.56\linewidth]{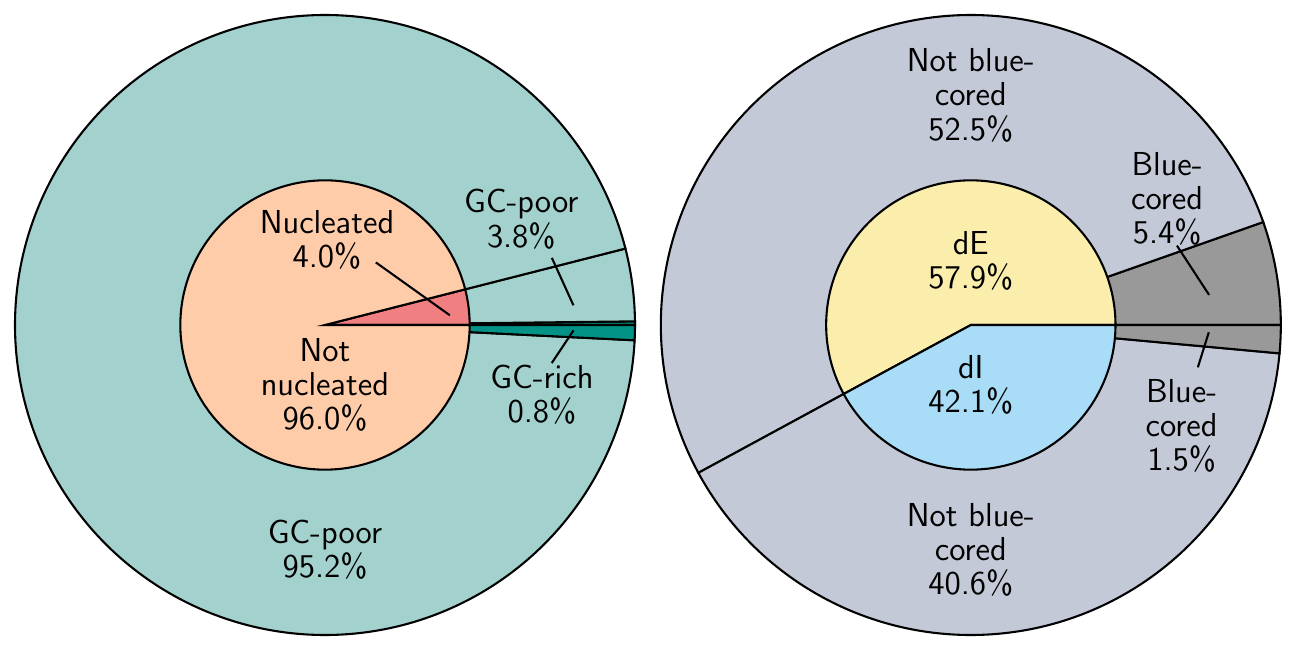}}
\caption{Morphological classification of the 2674 dwarf galaxy candidates. Darker shades on the outer circle correspond to the presence of the feature of interest.
\textit{Top left}: Nucleated fraction (outer circle) as a function of morphology (inner circle). \textit{Top middle}: GC richness (outer circle) as a function of morphology (inner circle). \textit{Top right}: Signs of disturbed morphology (outer circle) as a function of morphology (inner circle). \textit{Bottom left}: GC richness (outer circle) as a function of the nucleated fraction (inner circle). \textit{Bottom right}: Blue-cored fraction (outer circle) as a function of the morphology fraction (inner circle). A total of 1548 (57.9\%) dwarf galaxy candidates are classified as dE (dI: 1125 or 42.1\%), 106 (4.0\%) are nucleated, 26 (1.0\%) are potentially GC-rich, 118 (4.4\%) have a disturbed morphology, and 183 (6.9\%) are blue-cored. It is important to note that the galaxies classified as disturbed were explicitly dEs; hence, there is no disturbed dI.
}
\label{fig:pie_chart_morphology}
\end{figure*}

\subsection{\label{sc:allQ1tiles} All Q1 tiles: automatic selection}

We used the dwarf candidate sample of reference tile 102158889 to determine the selection cut and to calibrate the photometry of the other tiles. We followed the general methodology outlined in \citet{Habas2020} to determine the best parameters to select potential dwarf candidates. For the MATLAS data set, the cleanest separation between the visually identified dwarfs and background galaxies was found in plots of average surface brightness, $\langle \mu_{g} \rangle$, versus apparent magnitude, $g$. In particular, they found less scatter using average surface brightnesses instead of the more common central surface brightness, $\mu_{g,0}$. Therefore, for each galaxy in the Q1 MER catalogues, the magnitude, \IE, was computed using the \texttt{FLUX\_SEGMENTATION} parameter and the average surface brightness, $\langle \mu_{\IE} \rangle$, was computed using both the \texttt{FLUX\_SEGMENTATION} and \texttt{SEGMENTATION\_AREA} parameters. We did not attempt to convert \IE to absolute magnitudes, since there is no preferential distance to assume in the Q1 tiles. 

It should be noted, however, that the placement of the dwarf candidates in the \Euclid diagnostic plot relative to the background galaxies is different from the relative positions observed in the equivalent MATLAS diagnostic diagram. The MATLAS dwarfs follow a slightly shallower slope than the background galaxies, whereas the visually identified \Euclid dwarf candidates follow a sequence parallel to the main locus of points. A possible explanation for this difference may be the data reduction pipeline, since the MATLAS images were processed with the \texttt{EXLIXIR-LSB} pipeline which was specifically designed to preserve low surface brightness signals. Thus, the rest of the dwarf selection procedure in this work was modified for the Q1 data set.

Although the $\langle \mu_{\IE} \rangle$ -- \IE selection criterion is among the best at separating dwarf candidates from massive galaxies, it still identifies a high fraction of contaminants. Many of these are background S0 galaxies, spirals with diffuse halos, diffraction spikes, and spurious detections. We imposed a series of cuts, based on the following MER parameters, to further clean the candidate catalogue and remove badly formatted entries:
\begin{itemize}
\item {\texttt{VIS\_DET}} $= 1$;
\item \texttt{FLAG\_Y} $= 0 $ or $= 2$;
\item \texttt{FLUX\_SEGMENTATION} $> 0$;
\item \texttt{SEGMENTATION\_AREA} $>0$;
\item \texttt{RIGHT\_ASCENSION} $>0$;
\item \texttt{DECLINATION} $>-100$.
\end{itemize}
We intentionally did not impose any conditions on the quantity \texttt{POINT\_LIKE\_PROB}. In principle, this can be used to distinguish galaxies from stars, similar to the \texttt{CLASS\_STAR} parameter in \texttt{SourceExtractor}, but applying a cut here removed a handful of visually identified objects of interest from the sample. Thus, there are a few stars remaining in the candidate list.  

We identified two issues while looking at the photometric parameters from the MER catalogues of all Q1 tiles. Firstly, there is a small shift in the photometry in some of the tiles; this was fixed by fitting the narrow locus of stars (\texttt{POINT\_LIKE\_PROB}\,$\geq 0.8$; pink points in Fig.\,\ref{fig:selection}) in the reference tile and using this as the photometric calibration point for all other tiles. The magnitude of the surface brightness shift across all 25 tiles ranges from $-0.198$ to $0.364$, with a median of $-0.008$. Secondly, some tiles that are at the border of the Q1 footprint have only partial coverage, which reduces the number of detected sources in the tile. For the ten tiles with very few sources (102018211, 102102020053, 102020542, 102022971, 102041033, 102047412, 102158579, 102159781) or no stars (102021021, 102160873), no shift was applied and the photometry will be assumed to match that of the reference tile.

\begin{figure*}[ht!]
\centerline{\includegraphics[width=0.8\linewidth]{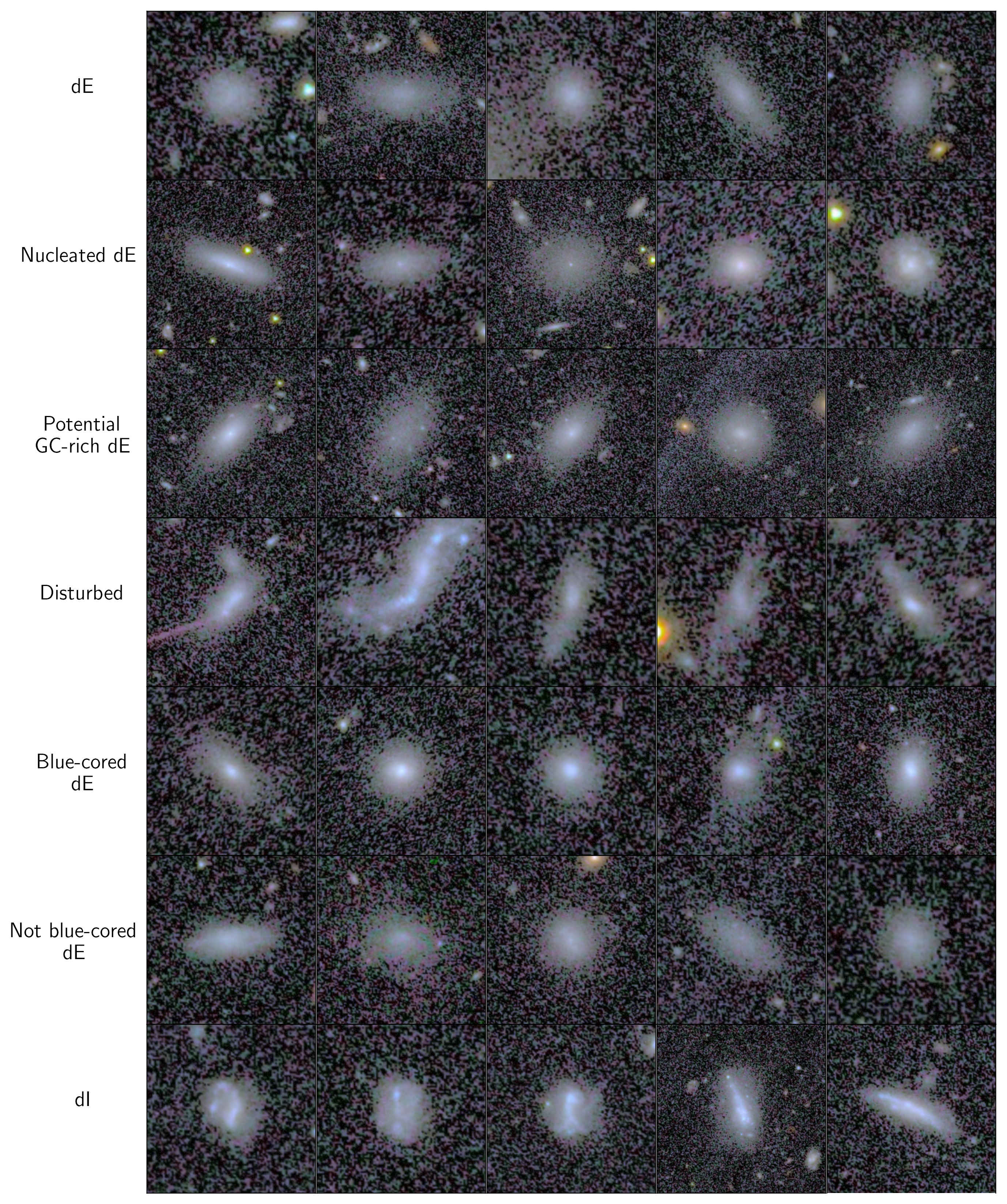}}
\caption{Cutouts of some dwarf candidates, taken from the VIS-NISP colour image created using the \IE\ band in blue, the \YE\ band in green, and the \HE\ band in red. The colours are projected onto the high-resolution \IE\ band to best reflect their appearance as detected. From top to bottom: dE; nucleated dE; potential GC-rich dE; disturbed morphologies; blue-cored dE; not blue-cored dE; and dI. The sizes of the cutouts are proportional to twice the area determined from the annotation of classifiers; north is up and east is to the left.}
\label{fig:cutouts-examples}
\end{figure*}

To determine an optimal selection cut, we matched the visual catalogue with the MER catalogue for the reference tile using a matching radius of 2\arcsec. Since the MER catalogue does not include detections within approximately \ang{;1;} of the tile edge -- these galaxies are, however, extracted from overlapping observations in neighbouring tiles -- we first removed these galaxies from our visual sample so that they would not register as missed detections. We then cleaned the visual-MER matched catalogue of problematic matches, such as: when the matching picked a nearby star/bright galaxy instead of the dwarf (which was not detected in the MER catalogue due to the proximity of the bright source); when the visual source was split into clumps in the MER catalogue; when the full extent of the diffuse dwarf was only detected in the MER catalogue as a very small source (\texttt{MER\_SEGMENTATION\_AREA} is too small compared to the visual extent of the dwarf leading to a wrong calculation of the surface brightness); when the source is an artefact; and when the source is near a stellar spike (and therefore the photometry may be incorrect). After this cleaning, we obtained a total of 682 dwarf candidates with MER photometry, making no distinction between different classification scores. We then used the most secure dwarf candidates with scores $\geq 0.8$ to produce isodensity contours in the surface brightness vs. magnitude plot for these robust 174 dwarf candidates (see Fig.\,\ref{fig:selection}). 

\begin{figure*}
\centerline{
\includegraphics[width=\linewidth]{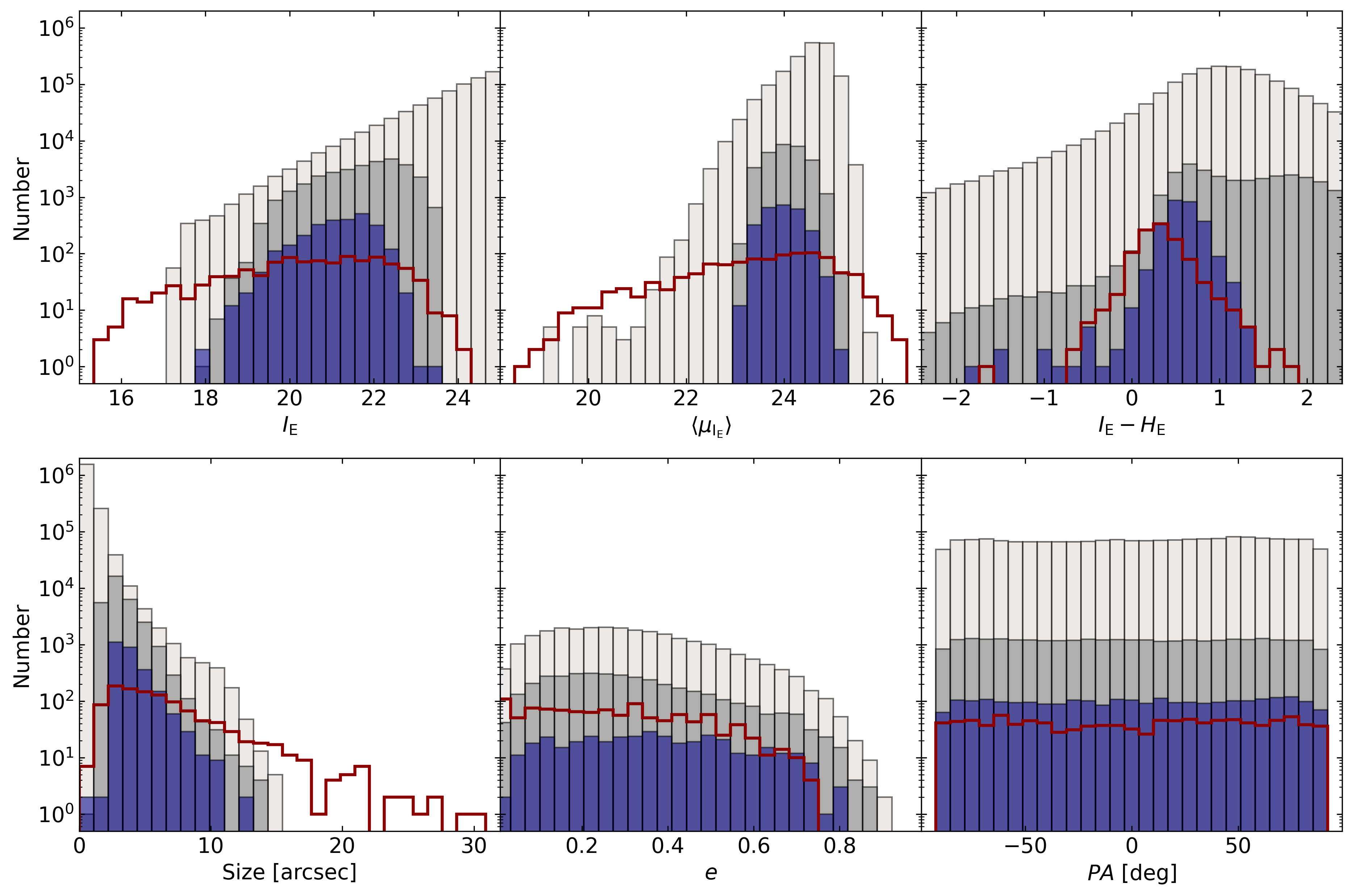}}
\caption{Distribution of the properties of our final dwarf candidates (blue-violet) overlaid on the properties of all galaxies in the MER catalogue (pale brown) and the automatically selected dwarf candidates in all 25 Q1 tiles (dark grey). The properties of the \Euclid ERO Perseus dwarfs from \citet{EROPerseusDGs} are shown for comparison (dark red). The Perseus dwarfs have colours and ellipticities similar to those of the Q1 dwarfs, but tend to be more spatially extended and with a wider range of surface brightness.}
\label{fig:params}
\end{figure*}

The outer contour of the isodensity plot was used to define an upper demarcation line between dwarf galaxies and likely background galaxies (see Fig.\,\ref{fig:selection}). Given the low statistics of dwarf candidates at the bright end of the contour, we manually extended the curve enough to capture the brightest candidates that would otherwise have been excluded by the line. With this adjustment, the selection cut recovers 97\% of the score $\geq 0.8$ dwarf candidates, 73\% of the score $\geq 0.5$ candidates, or 66\% of the cleaned candidates with no score cut. The position of this line was fixed and then applied to all Q1 tiles, producing a catalogue of \num{373745} sources.

Although the region of the selection plot that is populated by the cleaned sample is not as dense as the main locus of background galaxies, small shifts in the selection cut can sharply increase the number of sources that are extracted for visual inspection. For example, if we apply a vertical offset of $-0.1$\magarc\ to the original selection cut, we would add an additional 778 galaxies from the reference tile alone. Such a shift would allow us to also retain a further 61 dwarf candidates (no score cut applied) from the cleaned sample, but this is a very low return rate for the amount of time it would take to classify the additional galaxies. Alternately, one could consider a selection cut where the turnover at the bright end of the outer density contour is ignored and the slope of the contour is simply extended to the bright end; this would select an additional 521 objects from the reference tile, but with zero additional dwarf candidates. 

\begin{figure}[ht!]
\centerline{\includegraphics[width=0.95\linewidth]{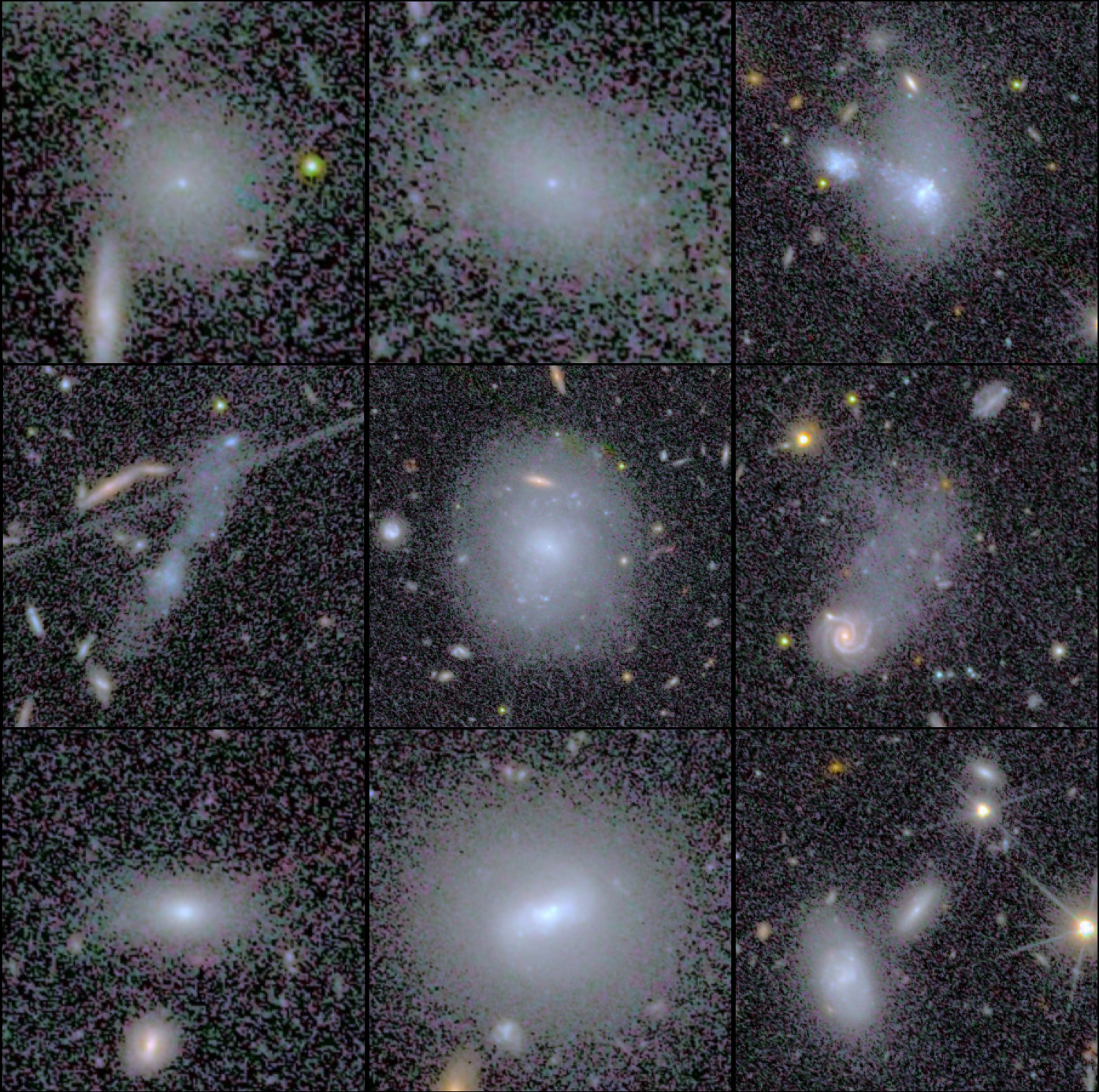}}
\caption{Examples of dwarf galaxy candidates of different types. From top left to bottom right: a nucleated dE with an elongated substructure connecting to the nucleus, possibly a signature of a GC migration and merging with the NSC; a traditional nucleated dE; a dI with clumps of star formation extending beyond the diffuse component; a pair of interacting dwarfs; a dI with a diffuse blue centre; a dI in superposition with a more distant spiral galaxy (having a spectroscopic redshift of $z=0.388$); a blue-cored dE; a dE with a spectroscopic redshift of $z=0.0272$ (or a distance of about 120\,Mpc) showing a double nucleus with inner `S-shape' features, likely associated with a NSC-NSC merger event; and a group of three dwarfs situated within a \ang{;;30} radius. The images are approximately \ang{;;11} $\times$ \ang{;;11} in size, except the image in the middle which is \ang{;2;} $\times$ \ang{;2;}, with north is up and east to the left.}
\label{fig:examples}
\end{figure}

\begin{figure}[ht!]
\centerline{\includegraphics[width=0.95\linewidth]{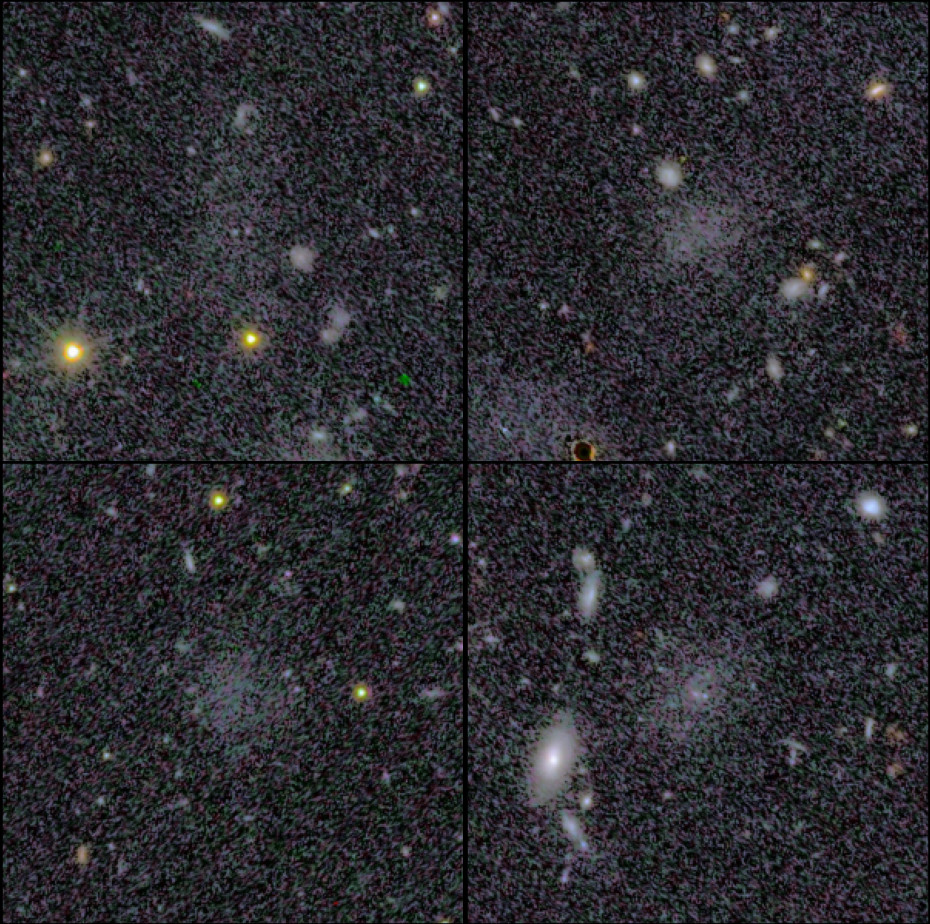}}
\caption{Examples of dwarf galaxy candidates with very low surface brightness that were identified during the visual inspection of the Q1 tiles. Three of these galaxies (top left, top right and bottom right) have a MER detection within a 2\arcsec\ matching radius, but are not retained by our automatic selection cut. The images are \ang{;1;} $\times$ \ang{;1;} in size, with north is up and east to the left.}
\label{fig:examplesfaint}
\end{figure}

\begin{figure*}
\centerline{
\includegraphics[width=\linewidth]{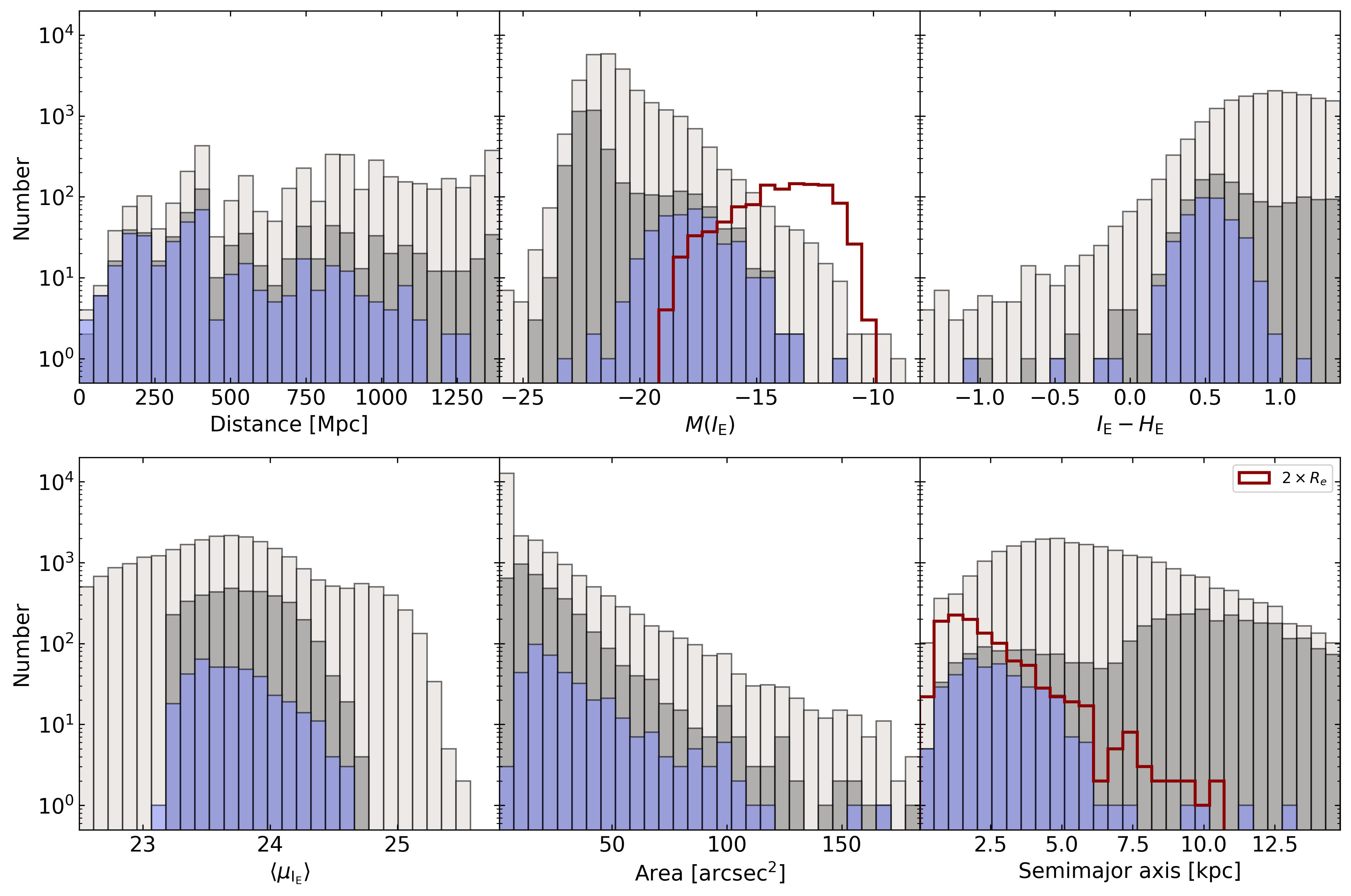}}
\caption{Distribution of the properties of the Q1 dwarf candidates with spectroscopic redshift (pale blue-violet) overlaid on the properties of all galaxies in the MER catalogue (pale brown) and the automatically selected dwarf candidates with spectroscopic redshift in all 25 Q1 tiles (dark grey). The absolute magnitudes of the \Euclid ERO Perseus dwarfs from \citet{EROPerseusDGs}, at a distance of 72\,Mpc, are shown for comparison (dark red). The Perseus dwarfs are closer than the majority of the dwarf candidates in the Q1 data and therefore are detected to fainter absolute magnitudes. The physical sizes of the Q1 and Perseus dwarf candidates (dark red) cover a similar range, assuming that the Perseus dwarfs have a semimajor axis approximately equal to twice their effective radius ($2\,\ \times\,R_{\rm e}$).}
\label{fig:specz}
\end{figure*}

Our selected objects were then matched with a morphological classification produced by \texttt{Zoobot} and based on the \Euclid Zooniverse project \citep{Q1-SP047} in order to clean the catalogue of as many background spiral galaxies as possible. The catalogue contains predictions for galaxies with an area $\geq 700$\,pixel$^2$ and of any magnitude. To identify the (background) spiral galaxies to remove from our sample, we selected the keywords (with format: \texttt{\{QUESTION\}\_\{ANSWER\}\_FRACTION}): \texttt{SMOOTH-OR-FEATURED\_FEATURED-OR-DISK\_FRACTION}\,$>$\,0.5 and \texttt{DISK-EDGE-ON\_NO\_FRACTION}\,$>$\,0.5 and \texttt{HAS-SPIRAL-ARMS\_YES\_FRACTION}\,$>$\,0.5. We visually checked some of the sources that were identified as spirals in the reference tile and confirmed their spiral nature. This step removed \num{51810} (14\%) background spiral galaxies, leaving a catalogue of \num{321934} sources.

We then examined the colours of the galaxies in the reference tile to remove any red ellipticals, a major source of contamination. We visually identified a sample of 353 red elliptical galaxies, computed their $\IE - \HE$ colours, and compared these values against the colours of the visually identified dwarf candidates. These colours were calculated using the MER parameters \texttt{FLUX\_VIS\_2FWHM\_APER} and \texttt{FLUX\_H\_2FWHM\_APER}, and corrected for Galactic extinction (as described in Sect.~\ref{sc:cat}). The colour distributions of these populations are shown in Fig.\,\ref{fig:colorcut}. We adopted a colour cut of $\IE - \HE < 1.4$, hence removing \num{155101} (48\%) red elliptical contaminant sources, and leaving \num{166834} candidates for all Q1 tiles. For the 25 tiles analysed in this paper, the catalogue after the automatic selection and cleaning contains a total of \num{14369} sources.

\subsection{\label{sc:allQ1tilesclass} All Q1 tiles: visual classification}

The \num{14369} dwarf galaxy candidates in the automatically produced catalogue for the 25 tiles were validated using the same online interface as described in Sect.~\ref{sc:visclass}. The classes were the same as for the reference tile, although we added two additional flags: `blue nucleus'; and `object of interest'. Both of these options could be selected independently of any other inputs.

To most efficiently classify the sample, we split the catalogue of dwarf candidates into four groups, with each group being examined by a subset of three classifiers. The classifiers for each group were selected after examining classification trends between individuals in the reference tile (e.g., who was more conservative or more liberal in the galaxies they classified as dwarfs), and we avoided grouping those on either end of the spectrum who classified too similarly, to avoid unbalanced voting between groups. The catalogue of dwarf candidates was first sorted by tile, then magnitude -- ideally, to increase our consistency with similar objects viewed one after another -- then evenly split between each of the four groups. The summed scores for objects within each group are shown in Fig.\,\ref{fig:group_classifications}. There are still some inconsistencies between groups, particularly in the use of the `unsure dwarf' classifications (this is most prominent in the summed scores $< 1.5$), but the distribution of votes above the median score seems reasonable. It should also be noted that the on-sky regions covered by groups three and four contain fewer massive galaxies (see Sect.~\ref{sc:results}) -- and presumably fewer dEs -- which could naturally have led to more `unsure` votes. 

For the remainder of this paper, the dwarf catalogue consists of the dwarf candidates with summed scores $\geq 2$, i.e., normalised scores $\geq0.67$. These are galaxies with at least (a) 1 `dwarf' and 2 `unsure dwarf' votes or (b) 2 `dwarf' votes. This returns 1136, 1156, 885, and 1648 dwarfs from each of the four groups, respectively. The apparent discrepancy between Group\,4 and the other three occurs amongst the smallest candidates, which Group\,4 tended to more confidently classify as dwarfs. To further balance systematics between the groups, we applied an area cut, as described below. Note that, by chance, this is also the size below which \citet{Q1-SP047} deemed the galaxy modelling to be unreliable. 

To explore the impact of area on our classifications, we examined the fraction of dwarf candidates per area bin in Fig.\,\ref{fig:sizecut}. For clarity, we merged the four groups into a single sample, but note that the fraction of dwarf galaxies in the smallest area bins are largely driven by a single group. Regardless if we consider our most secure candidates (score\,$=$\,1) or a more relaxed sample (score\,$\geq0.67$), the shape of the distributions are similar; we voted to keep a smaller fraction of candidates in the smallest area bins and higher fractions in the larger area bins (but there is significant contamination in every bin). For the most secure candidates in the sample, the fraction of dwarf candidates drops below 10\% at roughly 700\,pixel$^2$. In Fig.\,\ref{fig:dropouts}, it can be seen that there are legitimate dwarf candidates even at these small sizes, but it is increasingly difficult to identify structures in the smallest candidates that would distinguish dwarfs from background galaxies. Given the high levels of contamination at small sizes, the difficultly in consistently classifying the smallest dwarf candidates, and the need to normalise classifications between the groups, we adopted the 700\,pixel$^2$ size cut.  

After this size cut was applied, the final catalogue was then produced by applying a final score cut of 0.67 (2/3). The candidate list was subsequently inspected using \texttt{Jafar} to remove duplicates. The nine dwarf candidates with colours bluer than the main distribution, i.e., $\IE - \HE < -1.4$ were also visually inspected; four were background contaminants and were removed, while five were kept as they appear to be genuine dwarf candidates. However, of these five, three have colour image defects, making their colour measurements (and photometry) unreliable. These are flagged in Table\,\ref{appendix:struct-param1} and \ref{appendix:struct-param2}. The remaining two have colours $\IE - \HE = -1.8$ and $-1.4$. The final catalogue contains a total of 2674 dwarf galaxy candidates. 

\section{\label{sc:results} Results}

In total, 2674 dwarf candidates were identified in these 25 tiles, encompassing 14.25\,deg$^{2}$ of the sky, which translates to 188 dwarfs per deg$^{2}$. Assuming these 25 tiles are representative of the full region imaged by the Q1 EDFs and the EWS, we can extrapolate this number to estimate the presence of approximately \num{9900}, \num{470000} and 2.6 million dwarf candidates in the Q1 EDFs (53\,deg$^{2}$), DR1 (\num{2500}\,deg$^{2}$), and the EWS (\num{14000}\,deg$^{2}$), respectively. 

\subsection{\label{sc:morpho_gc_nsc} Morphological classification}

We examine the results of the visual classification of our dwarf candidates for the following classes: morphology; GC richness; presence of one or multiple nuclei; and presence of one of more blue central clumps. 

The breakdown in the number of dwarf candidates in each of the morphological classes is given in Fig.\,\ref{fig:pie_chart_morphology} and listed in Table\,\ref{appendix:visual-morph}. A total of 57.9\% were classified as dE, of which a small fraction (4.4\%) were identified as being a dE with disturbed morphology (possibly the result of tidal interaction, tidal tail formation or dwarf-dwarf merger). Very few (1.0\%) dwarfs were found to have two or more GC candidates. As presented in Sect.\,\ref{sc:speczdwarfs} below, our dwarf candidates span larger distances and therefore it is not surprising that this fraction is much lower than what was found for the more nearby dwarfs in the Perseus cluster (at a distance of 72\,Mpc), where 26\% of the dwarfs were visually classified as GC-rich (see a more detailed discussion in Sect.\,\ref{sc:GC-rich}). 
Likewise, there are few (4.0\%) nucleated dwarfs, and a small fraction (6.9\%) of candidates show a blue core. It should be noted that if a dwarf was classified as having both a blue core and a nucleus, it was not considered as having a nucleus. The blue core dwarfs show a wide range of visual structural properties, including unresolved blue cores, extended blue central clumps, complex and S-shape substructures, and finally double nuclei and multiple clumps. Examples of dwarf candidates in each of the visual morphological classes are shown in Fig.\,\ref{fig:cutouts-examples}.

\begin{figure}[ht!]
\includegraphics[width=\linewidth]{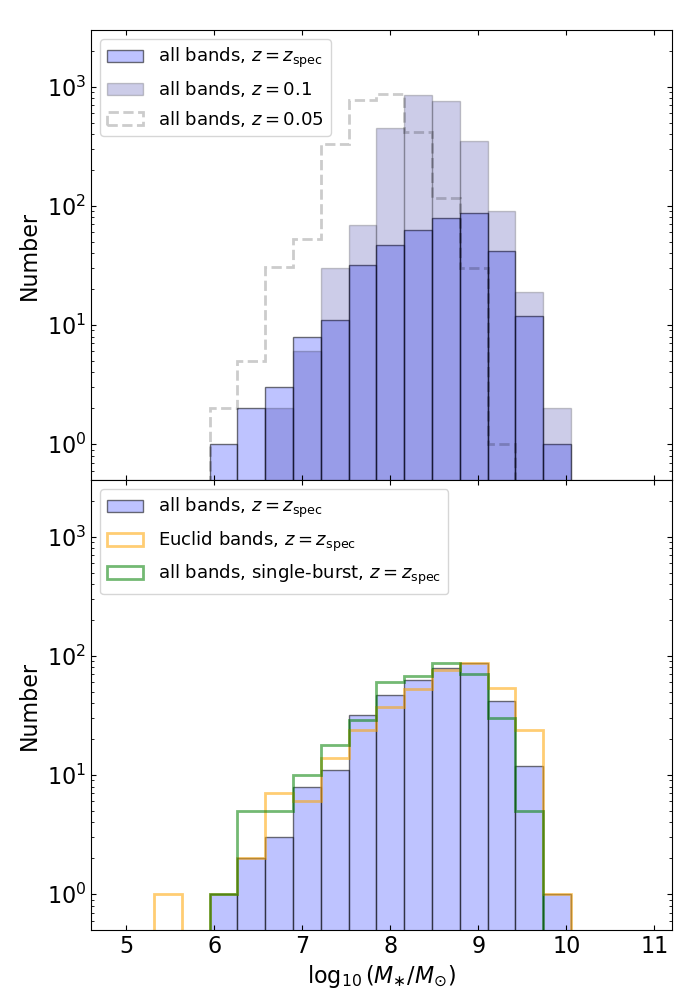}
\caption{Distribution of stellar masses for the Q1 dwarf candidates. \textit{Top:} Stellar mass distribution obtained using the reference SED fitting setup and all available bands for all dwarf galaxies, assuming $z=0.1$ (pale blue-violet) and $z=0.05$ (dashed blue-violet), as well as for those with spectroscopic redshifts $z_{\rm spec}$ (opaque blue-violet). \textit{Bottom:} Only dwarf galaxies with spectroscopic redshifts $z_{\rm spec}$, comparing the reference SED fitting results with those obtained using \Euclid bands alone (orange). Additionally, it includes results from an alternative SED-fitting approach that employs single-burst models with a broader range of metallicities (green).}
\label{fig:hist_masses}
\end{figure}

\begin{figure*}[ht!]
\centerline{
\includegraphics[width=1.0\linewidth]{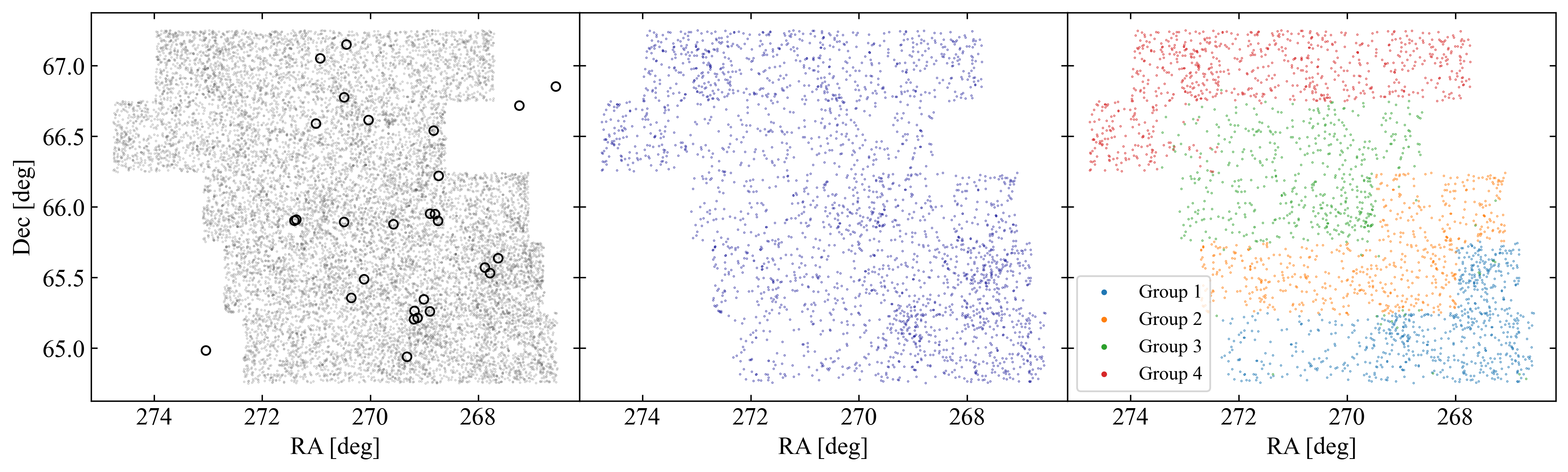}}
\caption{\textit{Left:} Spatial distribution of all candidates in the 25 tiles. The open black circles correspond to the positions of galaxies within a distance of about 200\,Mpc from the HECATE sample. It can be seen that several of these positions have corresponding overdensities in the distribution of the automatically selected dwarf candidates as well. \textit{Middle:} Spatial distribution of the final dwarf candidates where it can be seen that some of the overdensities around the massive galaxies shown in the left plot remain present in the final dwarf sample. \textit{Right:} Spatial distribution of the final dwarf candidates, colour coded by group. There are no large discontinuities in the number of dwarf candidates near the boundaries between groups. }
\label{fig:spatdist}
\end{figure*}

\subsection{\label{sc:properties} Photometric properties of the candidates}

Because our dwarf candidates were originally identified based on the sources detected and listed in the MER catalogue, we examined the range of some of the most important properties measured by the MER pipeline (e.g., \texttt{ISO\_MAG}, \texttt{SEGMENTATION\_AREA}, \texttt{SEMIMAJOR\_AXIS}, \texttt{ELLIPTICITY}, \texttt{POSITION\_ANGLE}) and compared them to both our automatically selected catalogue for all Q1 tiles (hence prior to visual classification) and all galaxies in the MER catalogue. The distribution of some of these properties are shown in Fig.\,\ref{fig:params}. The MER properties for the 2674 dwarf candidates can be found in Table\,\ref{appendix:struct-param1} and \ref{appendix:struct-param2}. 

As can be seen in Fig.\,\ref{fig:params}, the dwarf candidates in the final sample are rounder than the automatically selected dwarf candidates. All other parameters appear to be consistent with our expectations. Compared to the \Euclid ERO Perseus dwarfs \citep{EROPerseusDGs}, the Q1 dwarfs have the same range of colours and ellipticity, but tend to have smaller angular sizes and a wider range of surface brightnesses.

To highlight the diversity of the galaxies found in the Q1 fields, we show special examples of dwarf candidates in Fig.\,\ref{fig:examples}. This includes: (i) a nucleated dE with an elongated substructure connecting to the nucleus, likely a signature of a GC migration and merging with the NSC \citep{Poulain2025}; (ii) a traditional nucleated dE, similar to the nucleated dEs found in the Perseus cluster; (iii) a dI with clumps of star formation extending beyond the diffuse component, possibly triggered by an interaction; (iv) a pair of interacting dwarfs; (v) a dI with a diffuse blue centre; (vi) a dI in superposition with a more distant spiral galaxy (having a spectroscopic redshift of $z=0.388$); (vii) a blue-cored dE; (viii) a dE with a spectroscopic redshift of $z_=0.0272$ (or a distance of about 120\,Mpc) showing a double nucleus with inner "S-shape" features, likely associated with a NSC-NSC merger event \citep{Poulain2025}; and (ix) a group of three dwarfs situated within a \ang{;;30} radius. 

We also show in Fig.\,\ref{fig:examplesfaint} some examples of the dwarf candidates with the very low surface brightness that were visually identified in the Q1 images. Of these four examples, it is interesting to note that three have a MER detection, but are not retained by our automatic selection cut.

\subsection{\label{sc:speczdwarfs} Candidates with spectroscopic redshifts}

We have cross-matched our list of dwarf candidates with a spectroscopic catalogue from DESI \citep{Desi2024} using a 2\arcsec\ match radius. A total of 388 candidates (15\% of the total sample) were found to have a match; however no cuts in signal-to-noise ratio (S/N) have yet been applied because this information is not currently available. The distribution of the properties of the Q1 dwarf candidates with spectroscopic redshift is shown in Fig.\,\ref{fig:specz} and compared to the one of all galaxies in the MER catalogue and the automatically selected dwarf candidates with spectroscopic redshifts in the 25 Q1 tiles. The MER catalogue, automatically selected dwarf candidates, and final dwarf candidates have a median distance of 5448, 3848, and 404\,Mpc, respectively. The dwarf candidates with a spectroscopic redshift have the following properties: $\IE=17.8-22.5$, $\IE-\HE=-1.05-1.18$, $\langle \mu_{\mathrm{I_E}} \rangle=23.15-24.66$, \texttt{SEGMENTATION\_AREA}=$5.24-168.66$\,arcsec$^2$, and \texttt{SEMIMAJOR\_AXIS}=$0.67-2.92$\,arcsec. 

As can be seen in the top middle panel of Fig.\,\ref{fig:specz}, the Q1 dwarfs absolute magnitude distribution is shifted towards brighter values with respect to the ERO Perseus dwarfs. This difference is most likely due to the fact that the Perseus dwarfs are closer than the majority of the dwarf candidates in the Q1 data and therefore are detected to fainter absolute magnitudes. However, the difference could also be caused by a wrong spectroscopic redshift associated with some of the dwarf candidates; this would also explain the dwarf candidates with distances larger than about 500\,Mpc (Fig.\,\ref{fig:specz}, top left panel). Another reason could be that the faintest dwarf galaxies are missed due to our automatic selection cut, since the MER catalogue contains some sources at these faint magnitudes. It is also worth noting that the galaxies likely to have a spectroscopic redshift are presumably star-forming emission line objects that might have been detected at larger distances. The majority of our dwarfs are dEs that may have been selected at smaller distances. 

In terms of physical sizes, as shown in the bottom right panel of Fig.\,\ref{fig:specz}, the Q1 and Perseus dwarf candidates cover a similar range; this comparison however relies on the assumption that the Perseus dwarfs have a semimajor axis approximately equal to twice their effective radius ($2\,\ \times\,R_{\rm e}$).

\subsection{\label{sc:GC-rich} Potential GC-rich dwarf candidates}

With the spatial resolution and depth of the Q1 data, for dwarf galaxies we expect to detect GCs only out to about 100\,Mpc (\citealp{EP-Voggel}). At this distance threshold, we are only able to detect the brightest GCs. Considering that most dwarf galaxies in the EDF-N are at distances larger than 100\,Mpc, for typical dwarf galaxies with only a few GCs, we expect to see no GCs. However, in the case that some dwarf galaxies are indeed rich in GCs (more than 20), we expect that, given the shape of the GC luminosity function, a few GCs will be detected around those GC-rich dwarf galaxies. Therefore, in this work, the presence of two or more point-like sources (GC candidates) around dwarf galaxies are considered as a hint that the dwarf galaxy is potentially GC-rich. On the other hand, the lack of detection of GC candidates does not necessarily imply that the galaxy is GC-poor. In addition, since the majority of field dwarf galaxies are star-forming, any point-like source detections could instead be star-forming clumps within the galaxy, rather than GCs. In this work, we cautiously call dwarf galaxies with two or more apparent GC candidates `potential GC-rich' galaxies.

\subsection{\label{sc:masses} Stellar mass calculation}

To further characterise the dwarf sample and infer physically meaning interpretations, as well as constrain the stellar masses, we ran the software $\texttt{LePhare}$ (\citealt{Arnouts+99}; \citealt{Ilbert+06}), which performs a simple $\chi^{2}$ fitting between the stellar population synthesis theoretical models and the data. The multi-wavelength photometry was fitted using the models of \cite{Bruzual2003}, assuming: (i) a \cite{Chabrier01} initial mass function (IMF); (ii) the \cite{Calzetti2000} extinction curve; and (iii) an exponentially declining star-formation rate (SFR). The models allowed variations in star formation duration $(\tau \leq 30 \, \rm Gyr)$, metallicity $(0.2 \leq Z/Z_{\rm \sun} \leq 1)$, age $(\rm age \leq 14 \, Gyr)$, and internal extinction ($E(B-V) \leq 0.5$). This library favours high metallicity values and does not include metal-poor models that would better represent the stellar populations of certain dwarf galaxy types. To address this, we also performed SED fitting using single-burst models including metal-poorer metallicities in the range $0.005 \leq Z/Z_{\rm \sun} \leq 2.5$, $\rm age \leq 14$ Gyr and $E(B-V) \leq 0.5$. The spectrum was redshifted using a reference value for redshift (see below), or the spectroscopic value when available. Using the redshift, the distance is calculated to rescale the total luminosity and, therefore, properly scale the stellar mass.

We used magnitudes within an aperture of $2 \times$\,FWHM, $m_{\rm 2FWHM}$, and adopted two sets of magnitudes: Euclid bands only; and Euclid bands plus the ground-based optical bands, CFHT $u$, HSC $g$, CFHT $r$, Pan-STARRS $i$ (and related $1\sigma$ uncertainties). The photometry was corrected for Galactic extinction, as explained in Sect.~\ref{sc:cat}. In order to correct the stellar mass $M_{\ast}$ estimates from $\texttt{LePhare}$ for any missing flux, we used the total magnitudes, $m_{\rm S}$, derived from the S\'ersic fitting in the \IE\ band using:
\begin{equation}
\log_{10} (M_{\rm \ast}/M_{\odot}) = \log_{10} (M_{\rm \ast}/M_{\odot})^{\texttt{LePhare}} + 0.4 \times(m_{\rm 2FWHM}-m_{\rm S}),
\end{equation}

\noindent where $\log_{10} (M_{\rm \ast}/M_{\odot})^{\texttt{LePhare}}$ is the best-fit stellar mass estimate returned by $\texttt{LePhare}$. 

The spectroscopic redshift is available for only a fraction of the dwarf candidates, hence the distance is unknown for the rest of the sample. For the former, we obtained a stellar mass estimate using the spectroscopic redshifts. We also derived stellar mass estimates for the whole sample by adopting a redshift of $z=0.05$ and $z = 0.1$ (corresponding to a distance of approximately 200 and 400\,Mpc, respectively), as representative values for the true redshift distribution of the dwarf candidate sample. After removing objects with missing photometry and extinction corrections, we obtained stellar mass estimates for 2636 out of 2674 dwarfs. A total of 388 dwarf galaxies have a stellar mass estimate that was derived using their spectroscopic redshift. The average $1 \, \rm \sigma$ error on the stellar mass of the whole and spectroscopic sample are 0.15 and 0.05\,dex, respectively.

The results are shown in Fig. \ref{fig:hist_masses} and listed in Table\,\ref{appendix:mstar1} and \ref{appendix:mstar2}. In the top panel, we observe that the stellar mass distribution for the entire sample, obtained using the reference SED-fitting setup and assuming $z=0.1$, closely overlaps with that of the spectroscopic sample. Lowering the redshift assumption to $z=0.05$ has a minor effect on the stellar mass-to-light ratio, but significantly impacts the distance estimation, leading to a systematic shift in stellar mass normalization of about 0.6\,dex compared to the $z=0.1$ case. Focusing on the spectroscopic sample, we investigated the impact of using only \Euclid bands and adopting a different spectral library with single-burst models. We find that the resulting stellar masses remain robust against these changes, with negligible variations. 

The median, 5th, and 95th percentiles of the stellar mass distribution for the spectroscopic sample are $\log_{10} (M_{\rm \ast}/M_{\odot}) \simeq 8.6$, $7.3$, and $9.4$\,dex, with only a few galaxies in the $9.4$--$9.9$\,dex range and the least massive ones at $6$–$6.5$\,dex. Similar values are obtained when fitting the full sample assuming $z=0.1$. The stellar mass distribution of the dwarf candidates aligns with expectations and therefore confirms that our selection criteria were successful in identifying dwarf galaxies while minimising contamination from non-dwarfs.

\subsection{\label{sc:spatialdist} Two-dimensional spatial distribution} 

Mapping the spatial distribution of the identified dwarf galaxy candidates is essential for determining whether they are associated with specific host galaxies or exist as isolated objects. We selected adjacent tiles for this purpose. The spatial distribution of the 2674 identified dwarf candidates in the Q1 tiles is shown in Fig.\,\ref{fig:spatdist}. There are no large discontinuities in the number of dwarf candidates near the boundaries between groups. 

\begin{figure*}[ht!]
\centerline{
\includegraphics[width=0.9\linewidth]{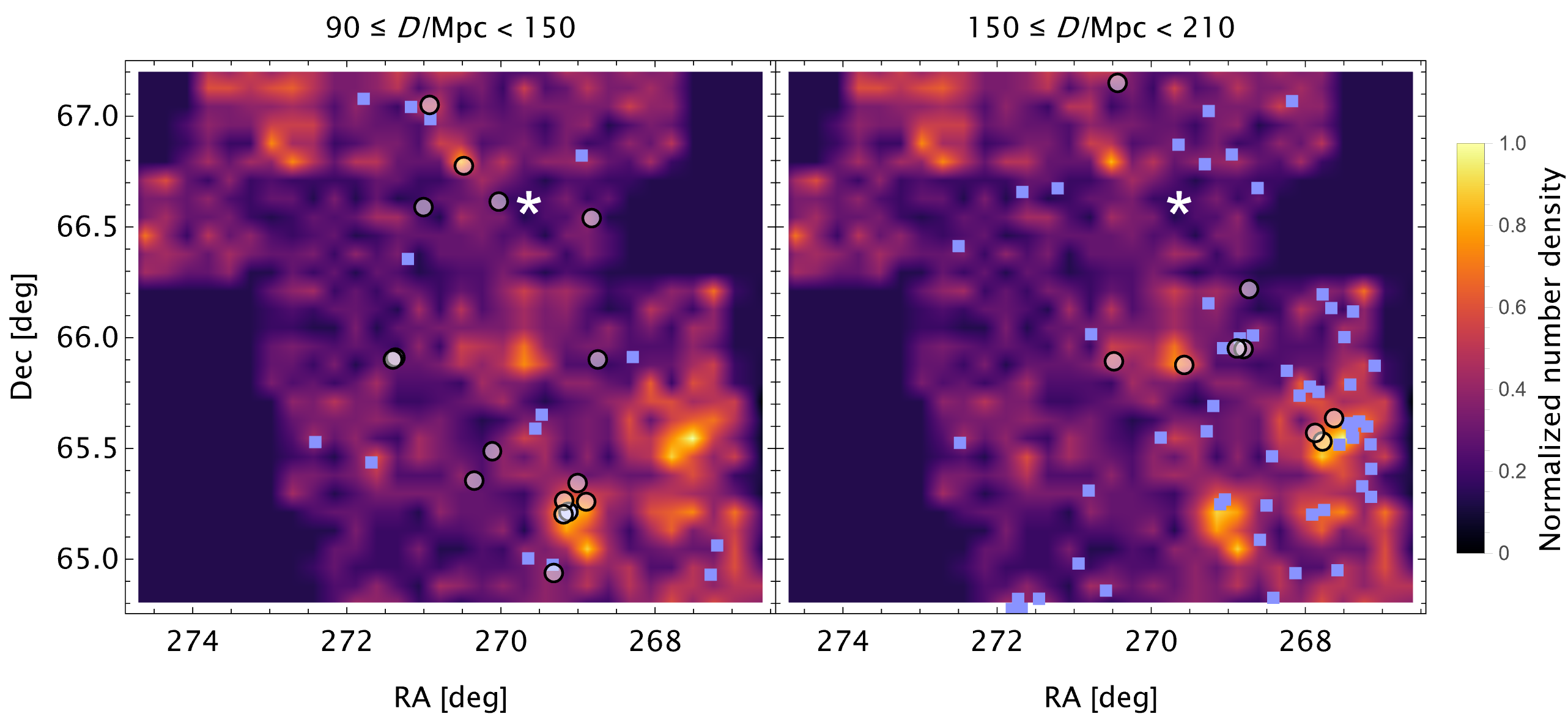}}
\caption{Spatial distribution of the final dwarf candidates is shown alongside the HECATE galaxies within the 25 tiles (marked as open black circles with a transparent white fill) and dwarf galaxies with measured spectroscopic redshifts (pale blue-violet filled squares). The colour map is based on the number of dwarfs within bins of RA and Dec, with this count normalised to a value of 1. According to the colour bar, orange-yellow colours indicate regions with dwarf galaxy overdensities. HECATE host galaxies and dwarfs with spectroscopic redshifts are displayed in the left and right panels, representing distances, $D$, in the ranges of $90 \leq D/{\rm Mpc} < 150$ and $150 \leq D/{\rm Mpc} < 210$, respectively. The white star indicates the location of the Cat's eye nebula where a lack of dwarfs is expected.}
\label{fig:soverdensities_with_zspecs}
\end{figure*}

To investigate their clustering with respect to possible host galaxies, we selected galaxies from The Heraklion Extragalactic CATaloguE\footnote{\url{https://hecate.ia.forth.gr/index.php}}
\citep[HECATE,][]{Kovlakas+21_HECATE}, which is a value-added catalogue of nearby galaxies, featuring robust distance measurements out to about 200\,Mpc. The black circles in Fig.\,\ref{fig:spatdist} correspond to the positions of HECATE galaxies, which are overlapping with the 25 tiles analysed in this paper. It can be seen that several of these positions have corresponding overdensities in the distribution of the host galaxies and the automatically selected dwarf candidates. 

To better quantify the distribution of the dwarf galaxies, Fig.\,\ref{fig:soverdensities_with_zspecs} displays a colour map, where yellow regions correspond to high dwarf overdensities and blue regions to lower overdensities. Additionally, dwarf galaxies with spectroscopic redshifts and HECATE galaxies, with distances, $D$, falling within two specific ranges, $90 \leq D/{\rm Mpc} < 150$ and $150 \leq D/{\rm Mpc} < 210$, are overplotted in the left and right panels, respectively\footnote{Distances for dwarfs with spectroscopic redshifts are calculated using a Hubble constant value of $H_{\rm 0} = 70\, \rm km \, s^{-1} \, Mpc^{-1}$.}.

In particular, we observe in the right panel of Fig.\,\ref{fig:soverdensities_with_zspecs} that the three HECATE galaxies located at $(\rm RA,\rm Dec)=$\,(\ang{267.9;;},\ang{65.6;;}) -- NGC\,6505, PGC\,2678657, and 2MASX\,J17503135+6538075 -- at a distance of around $180$--$185$\,Mpc, are not only overlapping with a clear overdensity in the dwarf sample, but are also associated with dozens of dwarfs with spectroscopic redshift and comparable distances. This confirms the presence of a clear overdensity of dwarf galaxies surrounding these three galaxies. The second most pronounced overdensity is found at $(\rm RA,\rm Dec)=$\,(\ang{269.0;;},\ang{65.2;;}), corresponding to the position of six host galaxies (PGC\,061115, PGC\,061122, PGC\,061140, PGC\,061148, PGC\,061160, NGC\,6536) with distances $D$\,$<$\,$150$\,Mpc (Fig.\,\ref{fig:soverdensities_with_zspecs}, left panel). However, we observe that no significant overdensity of dwarfs with similar distances is detected in this specific region. For the rest of the sky area, while some dwarf overdensities are found near potential host galaxies, there are not enough dwarf galaxies with spectroscopic redshifts to confirm a definitive association. No potential host galaxy from the HECATE sample is found at the location of the dwarf overdensity detected at $(\rm RA,\rm Dec)=$\,(\ang{273.0;;},\ang{66.9;;}). However, a visual inspection of the region reveals some extended galaxies that are not included in the HECATE sample and are therefore likely at larger distances.

\begin{figure*}[ht!]
\centerline{
\includegraphics[width=0.9\linewidth]{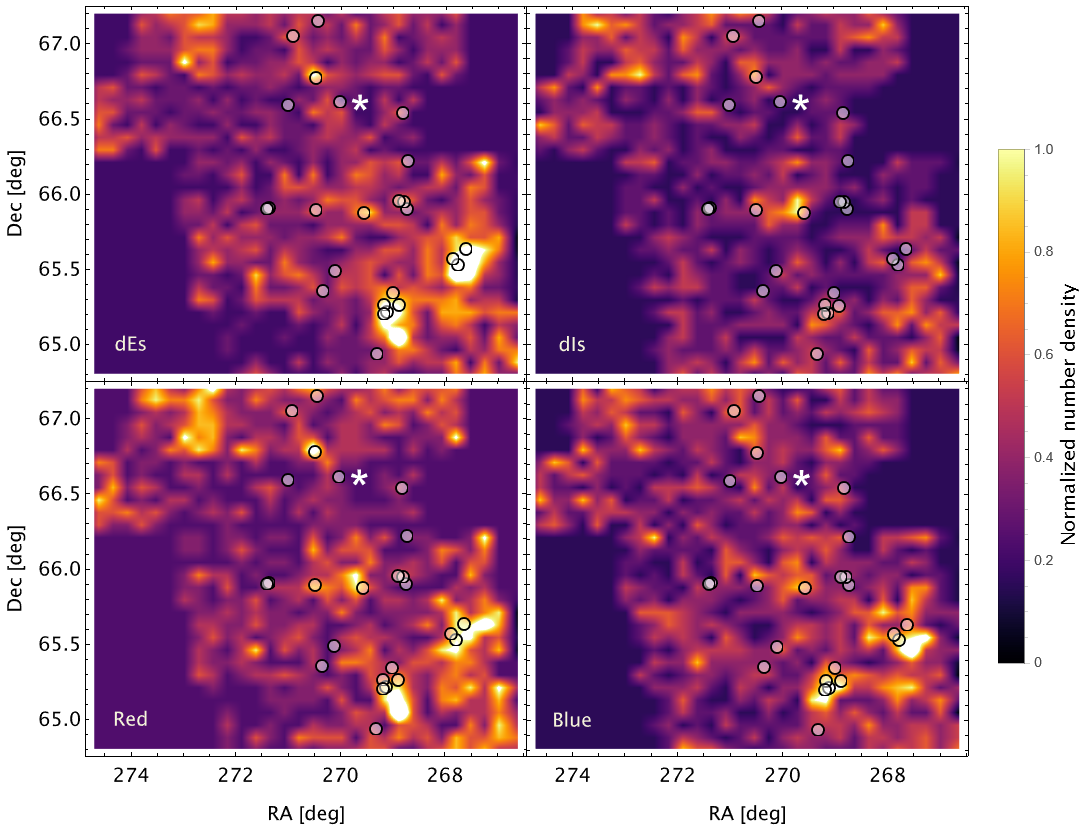}}
\caption{\textit{Top:} Spatial distributions of the final dwarf candidates, divided into the  dEs and dIs, are shown alongside the HECATE galaxies within the 25 tiles (marked as open black circles with a transparent white fill). \textit{Bottom:}: Same as top, but divided into the red ($I_{\rm E} - H_{\rm E} > 0.6$) and blue ($I_{\rm E} - H_{\rm E} \leq 0.6$) dwarf galaxies. The white star indicates the location of the Cat's eye nebula where a lack of dwarfs is expected.}
\label{fig:overdensities_types_colours}
\end{figure*}

Finally, it is interesting to explore how dwarf galaxies are distributed in terms of types and colour. Therefore, in Fig. \ref{fig:overdensities_types_colours} we show the colour map of the spatial distributions of dEs and dIs, and dwarfs divided in red and blue, defined using a threshold of $I_{\rm E} - H_{\rm E} = 0.6$. In particular we notice that the most prominent dwarf overdensities are dominated by dEs, while dwarfs identified as dIs are distributed more uniformly spread the field of view. In particular, dEs are predominant in the two dwarf overdensities in the bottom-right side of our field of view, where we have also seen that a large number of host galaxies are identified. Similarly the reddest dwarfs are on average more clustered than bluer dwarfs, the latter being, similarly to dIs, distributed more homogenously. However, in this last case, we also observe a clustering of bluer dwarfs around the overdensities in the bottom-right of our field of view. A more in-depth analysis of the field, which is beyond the scope of this paper, would be required to characterise the clustering of the sample and its variation with each single host.

\section{\label{sc:formation} Dwarf formation pathways}

Our analysis of the 25 tiles in the EDF-N has revealed different morphological classes of dwarf galaxies with a range of clustering properties. As discussed in Sect.~\ref{sc:spatialdist}, the dwarf candidates appear to be in some cases clustered around massive host galaxies, while in other parts of the sky, they are either clustered in regions empty of massive galaxies or found in isolation (or with one or two other dwarf companions).

\begin{figure}[ht!]
\centerline{
\includegraphics[width=0.45\linewidth]{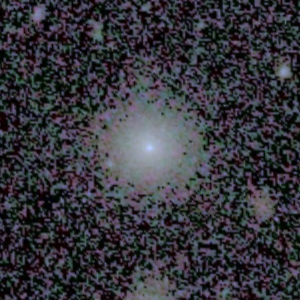}
\includegraphics[width=0.45\linewidth]{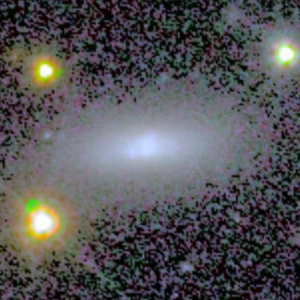}
}
\centerline{
\includegraphics[width=0.45\linewidth]{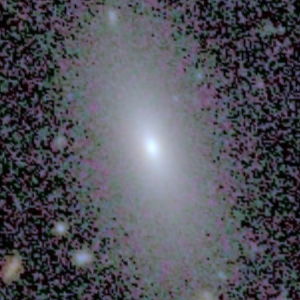}
\includegraphics[width=0.45\linewidth]{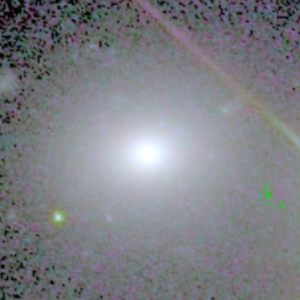}
}
\caption{Cutouts of some dwarf candidates, taken from the VIS-NISP colour image created using the \IE\ band in blue, the \YE\ band in green, and the \HE\ band in red. The colours are projected onto the high-resolution \IE\ band to best reflect their appearance as detected. \textit{Top left}: Example of a compact blue-cored dE. \textit{Top right}: dwarf with extended blue central clump. \textit{Bottom left} and \textit{bottom right}: candidates transitional dwarfs, one of which exhibiting an elongated central clump. The sizes of the cutouts are proportional to twice the area determined from the annotation of classifiers; north is up and east is to the left.}
\label{fig:bluecored2}
\end{figure}

A significant number of dEs exhibit a compact white or blue central nucleus, but only a very small number of them are GC-rich (0.8$\%$). In contrast, in the Perseus galaxy cluster we found that 24.5$\%$ of the dEs are GC-rich \citep{EROPerseusDGs}. This large discrepancy could be due to the ERO fields having a higher signal-to-noise ratio, better accuracy compared to the standard observing sequence used in the Q1 observations, or simply to the fact that more GC-rich dwarfs are found in denser environments \citep{EROPerseusOverview, Marleau2024}. Moreover, dwarf galaxies in the Perseus cluster are located at a distance of 72\,Mpc, which is below \Euclid's GC detection distance (about 100 Mpc; \citealt{EP-Voggel}; see also discussion in Sect.\,\ref{sc:GC-rich}). Figure \ref{fig:bluecored2} shows examples of dEs with a blue compact core (top left), dEs with an extended blue central clump (top right), and dwarf galaxies with a large central bright core (bottom left and right). 

The central nuclei in dEs have typical sizes of a few to a few tens of parsecs with a mass of nearly 1\% of their hosts \citep{Cote2006,Rossa2006}, and data from the {\it Hubble} Space Telescope have revealed that dE nuclei are slightly bluer than the rest of their host galaxies \citep{Lotz2004,Cote2006}. Moreover, detailed stellar population studies from integrated light spectroscopy have shown that they are, on average, younger than the host galaxies' stellar halos \citep{Chilingarian2009,Paudel2011,Guerou2015,Kacharov2018}. While typical dEs with a compact white central core have old nuclei and no star formation, the blue-cored ones exhibit a central region with active star formation. This suggests that they may be early-type dwarf galaxies in the process of forming a central nucleus \citep{Paudel2020}.

The formation channels of the central core in dEs are under debate. It is not clear whether the dEs formed with central nuclei that co-evolved together or formed later during the host galaxy evolution. Three formation channels have been envisaged: (1) nuclear clusters form via mergers of multiple globular clusters accreted through dynamical friction \citep{Tremaine1975,Oh2000,McLaughlin2006,Boker2010,Poulain2025}; (2) nuclear clusters form in situ from gas channelled into the centres of galaxies \citep{Milosavljevic2004,Bekki2007}; and (3) a significant merger event triggering a starburst in which an NSC forms and which self-quenches the galaxy \citep{Gray2024}. The first formation scenario is motivated by the fact that there are similarities between massive GCs and NSCs in terms of stellar mass and luminosity \citep{McLaughlin1999,Cote2006,denBrok2014,Poulain2021,EROPerseusDGs}. \citet{Paudel2020} presented a study of a sample of dEs located on the outskirts of the Virgo cluster that possess star-forming blobs near the isophotal centre of the galaxy. The top panels of their Figure\,1 show some of the dEs found in their analysis with SDSS spatial resolution (\ang{;;1.5}). In Fig.\,\ref{fig:bluecored2}, we show similar objects but with a \Euclid spatial resolution (\IE, \ang{;;0.20}), where the central blue cores look more like extended blobs. Inspection of the figure shows that the central blue regions appear large and, in some cases, exhibit elongated shapes. In Sect.\,\ref{sc:visclass}, we referred to these galaxies as transitional dwarf galaxies.

As discussed in \citet{Paudel2020}, star formation in these galaxies is concentrated within the nuclei region and presents high emission line metallicity (12 + $\logten$(O/H) $>$ 8.4\,dex), with a median of 8.59. In contrast, the stellar population of the halo is old and has a mean metallicity of log$_{10}$($Z/Z_{\odot}$) $\simeq -0.8$ (5 times more metal-poor), consistent with a typical dE. One of the key points to fully understand the nature of these galaxies is the origin of the gas supply that fuels star formation in the central regions. In the case of dEs with a blue core and residing in the outskirts of a galaxy cluster, they could have acquired their gas during a migration process from the inner regions to the outskirts; however, the metallicity of the star-forming regions would be lower than that observed in the galaxies from \citet{Paudel2020}, which reside in the outskirts of the Virgo cluster. An alternative scenario would be recycled gas flowing into the galaxy centre and re-igniting star formation \citep{Boselli2008}.

The formation pathway of the dwarfs and their NSC likely depends on the density of their local environment. In the satellites of the MATLAS galaxies \citep{Poulain2025} and in the dwarf galaxies detected by \Euclid in the Perseus galaxy cluster \citep{EROPerseusDGs}, there is evidence of the GC-inspiral scenario supported by the complex structure in the nuclei of the dwarf galaxies. This formation scenario is also supported by the presence of a large percentage (24.5\%) of GC-rich dEs in the Perseus cluster. In our visual inspection of the 25 Q1 tiles, we also find evidence of such nucleus subtructure, as seen in Fig.\,\ref{fig:examples} (top left) and discussed in Sect.\,\ref{sc:properties}. 

One important aspect to consider is the fact that the some of the dEs with a large central blue or white core detected in Q1 are isolated dwarfs, and therefore alternative formation mechanisms of a central star-forming region in these galaxies must be envisaged. \citet{Gray2024} used a suite of cosmological high resolution zoom simulations named EDGE \citep{Agertz2020} to study an isolated dwarf galaxy with virial mass of around $5\times10^{9}\,M_{\odot}$ at redshift $z = 0$. The simulations show that dwarfs are quenched by reionisation, but retain a significant reservoir of gas that is unable to cool and form stars. Sometime after reionisation, the dwarfs then undergo a major merger that triggers rapid gas cooling, leading to a significant starburst. An NSC forms in this starburst that then quenches star formation thereafter. Two possibilities are predicted: (1) formation of a dense NSC in the dwarf that started its assembly history at later epochs; and (2) no formation of a central core in dwarfs that had an earlier assembly history. This means that isolated galaxies can form a dense central core via merging events, and it is likely that the dEs with star-forming blue or white cores found in the Q1 data are going through a process of NSC formation, while evolving in nucleated quenched dEs.

Other scenarios for the origin of dEs with large white or blue core involve: (1) the conversion of gas-rich disc galaxies into gas-poor early-type dwarfs; and (2) a primordial origin. The first involves gas removal, with subsequent tidal heating, harassment, and ram-pressure stripping \citep{Gunn1972,Moore1996,Lake1998}. The process of ram pressure stripping alone is envisaged by \citet{Boselli2008} and \citet{Steyrleithner2020}, while \citet{Chilingarian2009} suggested a process involving major/minor mergers. It is worth noting that the transformation of gas-rich late-type galaxies to early-type dwarfs would be more likely in crowded environments like galaxy groups or clusters, where a significant portion of these transitional dwarf galaxies is found \citep{Makarov2011}. However, some of the galaxies detected in Q1 exist in isolation, as shown in this analysis, and the scenario of a primordial formation would be more plausible. As shown in \citet{Ann2024}, such a scenario is challenged by the fact that the star formation histories (SFHs) of transitional dwarf galaxies exhibit an active period of more recent star formation, similar to the SFHs observed in dEs. The SFHs of the isolated dwarfs reveal two distinct periods of active star formation. The first period ended about 6\,Gyr ago and witnessed the formation of predominantly old, metal-poor stars ($Z = 0.0004$), while the second period, which ended around 1 Gyr ago, contributed to the formation of moderately old stellar populations with intermediate metallicity. Moreover, the SFHs of transitional dwarf galaxies show a clear dependence on stellar mass, with more massive dwarfs forming stars at earlier epochs. From these findings, it seems that the scenario of primordial formation is unlikely.

As shown in the top panel of Fig.\,\ref{fig:overdensities_types_colours}, early-type dwarfs tend to cluster around massive galaxies and, therefore, the formation of a central star-forming region, and the NSC, is likely affected by the high-density environment. In the case of dEs with no obvious massive galaxy host, the more plausible scenario would be an in-situ formation, where the central core formed from the gas channelled into the centre of the dwarf, in certain instances possibly triggered by a dwarf-dwarf merger event. Future in-depth investigations of the stellar populations in these galaxies are already planned and will help to shed light on the formation and evolution of dEs in different environments.  

\section{\label{sc:conclusion} Conclusion}

We have taken advantage of the unprecedented depth, spatial resolution, and field of view of the \cite{Q1cite}, consisting of a total of 351 tiles located in the Euclid Deep Fields (EDF-N, EDF-S, EDF-F) and the Lynds Dark Nebula LDN1641, to build a census of dwarf galaxies in the 203 tiles with good quality data and full coverage. A representative sample of 25 contiguous tiles in the EDF-N were selected, covering a total area of 14.25 deg$^{2}$, for a first exploration of the Q1 data set. The analysis of the remaining tiles will be presented in a future paper.

The dwarf galaxy candidates were identified using a semi-automatic detection method, based on properties measured by the \Euclid pipeline and released as part of the OU-MER catalogue. A detailed visual analysis of a reference tile was used to calibrate the automatic selection method for all Q1 fields. A selection cut in surface brightness and magnitude was used to produce an initial dwarf candidate catalogue, followed by a cut in morphology (removing background spiral galaxies) and $\IE - \HE$ colour (removing red ellipticals). This catalogue was then visually classified to produce a final sample of dwarf candidates as well as to determine the morphology (dE, dI, disturbed), number of nuclei, potential GC richness, and the presence of a blue core. The distribution of the most relevant photometric and morphological parameters contained in the MER catalogue for each of the dwarf candidate was then obtained for this first sample of Q1 dwarfs. Some of our key findings are highlighted below.

\begin{itemize}

\item After applying a score and size cut, a total of 2674 dwarf candidates were identified over a region of 25 tiles, therefore covering 14.25 deg$^{2}$. In terms of number per sky area, this corresponds to 188 dwarfs per deg$^{2}$. Extrapolating this number to all Q1 Euclid Deep Field regions (53\,deg$^{2}$), the sky area covered by DR1 (\num{2500}\,deg$^{2}$), and the Euclid Wide Survey (\num{14000}\,deg$^{2}$), we calculate approximately \num{9900}, \num{470000} and 2.6 million dwarf candidate galaxies, respectively.
\\
\item In the final dwarf sample, 57.9\% were classified as dE, of which a small fraction (4.4\%) were identified as being a dE with disturbed morphology (possibly the result of tidal interaction, tidal tail formation, or dwarf-dwarf merger). Very few (1.0\%) dwarfs were found to have two or more GC candidates, compared to 26\% for the dwarfs in the Perseus cluster. This is possibly due to the fact that the Q1 dwarfs span a larger range of distances and probe different large-scale environments. Likewise, there are few (4.0\%) nucleated dwarfs, and a small fraction (6.9\%) of candidates show a blue core. These show a wide range of visual structural properties, including unresolved blue cores, extended blue central clumps, complex and S-shape substructures, and finally double nuclei and multiple clumps.
\\
\item The properties of the final dwarf candidates, as compared to the automatically selected dwarf candidate catalogue, appear to be rounder. Compared to the \Euclid ERO Perseus dwarfs, the Q1 dwarfs have the same range of colours and ellipticities, but tend to have smaller angular sizes and a wider range of surface brightnesses.
\\
\item We have cross-matched our list of dwarf candidates with a spectroscopic catalogue from DESI \citep{Desi2024} using a 2\arcsec\ match radius. A total of 388 candidates (15\% of the total sample) were found to have a match. Based on this spectroscopic sample, we find the median of the distance distribution for the dwarf candidates to be around 400\,Mpc. 
\\
\item The Q1 dwarfs with matched spectroscopic redshifts have a brighter range of absolute magnitudes than the Perseus dwarfs, most likely due to the fact that the majority of the Q1 dwarfs are found at greater distances. The physical sizes of the Q1 and Perseus dwarfs cover a similar range, assuming that the Perseus dwarfs have a semimajor axis approximately equal to twice their effective radius.
\\
\item Using SED fitting, the stellar mass range for the spectroscopic sample is $\log M_{\ast}/M_{\odot} \sim 7.3$--$9.4$, with only a few galaxies with larger masses. This implies that our selection criteria were successful in identifying dwarf galaxies while minimising contamination by other sources.
\\
\item Several of the positions of the bright (massive) galaxies have corresponding overdensities in the final dwarf sample. The most prominent dwarf overdensities are dominated by dEs, while dwarfs identified as dIs are distributed more homogeneously across the field of view.
\\
\item A significant number of Q1 dEs exhibit a compact white or blue central nucleus. Since early-type dwarfs tend to cluster around massive galaxies, the formation of a central star-forming region, and the NSC, is likely affected by the high density environment. In the case of dEs with no obvious massive galaxy host, the more plausible scenario would be an in-situ formation, where the central core formed from the gas channelled into the centre of the dwarf.
\end{itemize}

The Q1 images have demonstrated the capability of \Euclid to detect and characterise dwarf galaxies across a range of distances, stellar masses and environments. Due to the large range of distances probed by the Q1 data, contamination by background galaxies remains a major challenge. Future work will make use of the results presented in this paper to study in more detail the properties of these dwarfs and optimise their detectability.

\begin{acknowledgements}
\AckQone
\,\AckEC

\,This work used data products based on data from UNIONS, a scientific collaboration using three Hawaii-based telescopes: CFHT, Pan-STARRS, and Subaru \url{www.skysurvey.cc}\,. This research utilised data from the Dark Energy Spectroscopic Instrument (DESI). DESI is managed by the Lawrence Berkeley National Laboratory and is supported by the U.S.\ Department of Energy's Office of Science. The project is a collaboration involving numerous institutions and funding agencies worldwide. M.\,Poulain is supported by the Academy of Finland grant number 347089. A.\,Ferr\'e-Mateu acknowledges support from RYC2021-031099-I and PID2021-123313NA-I00 of MICIN/AEI/10.13039/501100011033/FEDER,UE,NextGenerationEU/PRT. R.\,Habas and M.\,Cantiello acknowledge funding from the Italian INAF Large Grant 12-2022. E. Sola is grateful to the Leverhulme Trust for funding under the grant number RPG-2021-205.  
This research has made use of the SIMBAD database, operated at CDS, Strasbourg, France.

\end{acknowledgements}

\bibliography{paper, Euclid, Q1}

\begin{appendix}
\onecolumn

\section{Properties of the dwarf galaxy candidates} 

\label{AppendixA}

The dwarf galaxy candidates are classified by visual inspection with the methodology described in Sect.~\ref{sc:method}. The final sample contains 2674 dwarf galaxies of which 57.9\% are classified as dE, and 42.1\% as dI. The dE galaxies are further classified as nucleated (4.0\%), blue-cored (6.9\%), potentially GC-rich (1.0\%), and with disturbed morphology (4.4\%).

Table\,\ref{appendix:visual-morph} lists the entire sample of dwarf galaxy candidates, ordered by increasing RA. The column definitions are as follows: a unique identifier (ID) based on the \texttt{OBJECT\_ID} parameter in the MER catalogue; RA in degrees; Dec in degrees; morphology (either dE or dI); GC-rich flag (recall that in the visual classification we defined dwarfs with $N_{\rm GC} \geq 2$\,GCs as GC-rich); a flag for the presence of a nucleus (Nucleated); a disturbed morphology flag (Disturbed); and a blue-cored flag (Blue-cored). Both dE and dI galaxies could be classified as disturbed.

Table\,\ref{appendix:struct-param1} lists the photometric and structural parameters of the dwarf galaxy candidates (ordered by increasing RA) obtained from the MER catalogue. The column definitions are as follows: a unique identifier (ID) based on the \texttt{OBJECT\_ID} parameter in the MER catalogue; RA in degrees; Dec in degrees; apparent magnitude in the \IE\ filter (\IE, computed using the \texttt{FLUX\_SEGMENTATION} parameter in the MER catalogue, see Sect.~\ref{sc:allQ1tiles}); Area in arcsec$^2$ (Area, see Sect.~\ref{sc:allQ1tiles}); ellipticity ($e$); position angle in degrees (PA); and average surface brightness in mag~arcsec$^{-2}$ ($\langle \mu_{\IE} \rangle$, see Sect.~\ref{sc:allQ1tiles}) ordered by increasing RA. The uncertainties reported in the tables are calculated using propagation of error formulae using the uncertainties reported in the MER catalogue. The three parameters for which we could not compute uncertainties are the area, ellipticity and position angle. The area is based on the \texttt{SEGMENTATION\_AREA} parameter in the MER catalogue which has no associated uncertainty in the catalogue. The ellipticity and position angle are based on the \texttt{POSITION\_ANGLE} and \texttt{ELLIPTICITY} parameters in the MER catalogue, respectively, which have the associated uncertainty parameters \texttt{POSITION\_ANGLE\_ERR} and \texttt{ELLIPTICITY\_ERR}; however, the respective column entries are empty.

Table\,\ref{appendix:struct-param2} lists the aperture magnitudes and extinction corrections (EC) described in Sect.~\ref{sc:cat}. The column definitions are: a unique identifier (ID) based on the \texttt{OBJECT\_ID} parameter in the MER catalogue; the apparent magnitude in the \IE\ filter ($I_{{\rm E,2FWHM}}$, computed using the \texttt{FLUX\_VIS\_2FWHM\_APER} parameter in the MER catalogue); the extinction correction in \IE; the apparent magnitude in the \YE\ filter ($Y_{{\rm E,2FWHM}}$, computed using the \texttt{FLUX\_Y\_2FWHM\_APER} parameter in the MER catalogue); the extinction correction in \YE; apparent magnitude in the \JE\ filter ($J_{{\rm E,2FWHM}}$, computed using the \texttt{FLUX\_J\_2FWHM\_APER} parameter in the MER catalogue); the extinction correction in \JE; the apparent magnitude in the \HE\ filter ($H_{{\rm E,2FWHM}}$, computed using the \texttt{FLUX\_H\_2FWHM\_APER} parameter in the MER catalogue); and the extinction in \HE. The dwarf galaxy candidates are ordered by increasing RA.

Table\,\ref{appendix:mstar1} and Table\,\ref{appendix:mstar2} list stellar mass estimates for the dwarf galaxy candidates as described in Sect.~\ref{sc:masses} and shown in Fig. \ref{fig:hist_masses}. Table\,\ref{appendix:mstar2} only lists dwarf galaxy candidates for which spectroscopic redshifts are available. The column definitions for Table\,\ref{appendix:mstar1} are: a unique identifier (ID) based on the \texttt{OBJECT\_ID} parameter in the MER catalogue; RA in degrees; Dec in degrees; the stellar mass estimate obtained via the reference SED fitting setup and all available photometric bands assuming $z=0.1$ ($M_{\ast,\mathrm{z}=0.1}$); and the stellar mass estimate using all bands assuming $z=0.5$ ($M_{\ast,z=0.05}$). The column definitions for Table\,\ref{appendix:mstar2} are: the unique identifier (ID); RA in degrees; Dec in degrees; the stellar mass estimate obtained via the reference SED fitting setup and all available photometric bands assuming $z=z_\mathrm{spec}$ ($M_{\ast,z=z_\mathrm{spec}}$); the stellar mass estimate obtained using Euclid bands alone ($M_{\ast,z=z_\mathrm{spec} (\mathrm{\Euclid\,bands})}$); and the stellar mass estimate obtained by employing a single-burst model in the SED fitting approach ($M_{\ast,z=z_\mathrm{spec} (\mathrm{single-burst})}$). The uncertainties given in both tables were estimated by using the 16th, 50th and 84th quantiles of the fitting routine and calculating the upper limit as (84th-quantile$-$50th-quantile) and the lower limit as (50th-quantile$-$16th-quantile). 

\begin{table*}[ht!]
\caption{Visual properties of the dwarf galaxy candidates.\tablefootmark{$\dagger$}}
\label{appendix:visual-morph}

\centering
\resizebox{\textwidth}{!}{
\begin{tabular}{cccccccc}
\hline
\hline
\noalign{\smallskip}
\omit\hfil ID \hfil & \omit\hfil RA \hfil & \omit\hfil Dec \hfil & \omit\hfil Morphology \hfil & \omit\hfil \phantom{00}GC-rich\phantom{00} \hfil & \omit\hfil Nucleated \hfil & \omit\hfil \phantom{00}Disturbed \hfil & \omit\hfil \phantom{00}Blue-cored \hfil\\
   & \omit\hfil [deg] \hfil & \omit\hfil [deg] \hfil & \omit\hfil & \omit\hfil & \omit\hfil & \omit\hfil & \omit\hfil\\
\hline
\noalign{\smallskip}
2665542189652466511 & 266.554219 & 65.246651 & dI & No & No & No & No \\
2665815442648786462 & 266.581544 & 64.878646 & dE & No & No & No & No \\
2665837571652492797 & 266.583757 & 65.249280 & dE & No & No & No & Yes \\
2665847551651610342 & 266.584755 & 65.161034 & dI & No & No & No & Yes \\
2665914298650060584 & 266.591430 & 65.006058 & dE & No & No & No & Yes \\
\vdots & \vdots & \vdots & \vdots & \vdots & \vdots & \vdots & \vdots\\
2747385300663221364 & 274.738530 & 66.322136 & dI & No & No & No & No \\
2747397180663791709 & 274.739718 & 66.379171 & dE & No & No & No & Yes \\
2747513424663839120 & 274.751342 & 66.383912 & dE & No & Yes & No & No \\
2747532158662611810 & 274.753216 & 66.261181 & dE & No & No & No & No \\
2747550219666016765 & 274.755022 & 66.601677 & dI & No & No & No & No \\
\hline 
\end{tabular}}
\tablefoot{
\tablefoottext{$\dagger$}{Table\,\ref{appendix:visual-morph} is only available in electronic form at the CDS via anonymous ftp to cdsarc.u-strasbg.fr (130.79.128.5) or via \url{http://cdsweb.u-strasbg.fr/cgi-bin/qcat?J/A+A/}.}
}
\end{table*}

\begin{table*}[hp!]

\small
\centering
\newcommand{\pd}{\phantom{1}}
\setlength{\tabcolsep}{3.25pt}
\caption{Photometric and structural properties of the dwarf galaxy candidates.\tablefootmark{$\dagger$}}
\label{appendix:struct-param1}
\smallskip
\smallskip

\resizebox{\textwidth}{!}{%
\begin{tabular}{cccccccc}
\hline
\hline

&&&&&&&\\[-7pt]
\omit\hfil ID \hfil & \omit\hfil RA \hfil & \omit\hfil Dec \hfil & \omit\hfil \IE\ \hfil & \omit\hfil Area \hfil & \omit\hfil $e$ \hfil & \omit\hfil PA \hfil & \omit\hfil $\langle \mu_{\IE} \rangle$ \hfil \\
\omit\hfil  & \omit\hfil [deg] \hfil & \omit\hfil [deg] \hfil & \omit\hfil [mag] \hfil & \omit\hfil [$\text{arcsec}^2$] \hfil & \omit\hfil & \omit\hfil [deg] \hfil & \omit\hfil [mag arcsec$^{-2}$] \hfil \\
&&&&&&&\\[-8pt]
\hline
&&&&&&&\\[-8pt]
2665542189652466511 & 266.554219 & 65.246651 & $21.00 \pm 0.07$ & 15.96	& 0.341045 & \phantom{$-$}56.254025 & $23.96 \pm 0.07$  \\
2665815442648786462 & 266.581544 & 64.878646 & $21.95 \pm 0.07$ & 10.82 & 0.173286 & \phantom{$-$}55.904274 & $24.48 \pm 0.07$ \\
2665837571652492797 & 266.583757 & 65.249280 & $21.54 \pm 0.06$ & \phantom{0}7.93 & 0.735860 & $-80.007446$ & $23.74 \pm 0.06$ \\
2665847551651610342 & 266.584755 & 65.161034 & $19.83 \pm 0.07$ & 33.10 & 0.365819 & \phantom{$-$}71.216850 & $23.58 \pm 0.07$ \\
2665914298650060584 & 266.591430 & 65.006058 & $21.08 \pm 0.08$ & 14.92 & 0.227967 & \phantom{$-$}61.733612 & $23.97 \pm 0.08$ \\
\vdots & \vdots & \vdots & \vdots & \vdots & \vdots & \vdots & \vdots \\
2747385300663221364 & 274.738530 & 66.322136 & $22.27 \pm 0.07$ & \phantom{0}8.44 & 0.264453 & $-72.742409$ & $24.59 \pm 0.07$ \\
2747397180663791709 & 274.739718 & 66.379171 & $19.59 \pm 0.06$ & 41.67 & 0.137528 & $-52.616344$ & $23.63 \pm 0.06$ \\
2747513424663839120 & 274.751342 & 66.383912 & $20.81 \pm 0.06$ & 13.17 & 0.046551 & \phantom{$-$}67.143562 & $23.60 \pm 0.06$ \\
2747532158662611810 & 274.753216 & 66.261181 & $21.87 \pm 0.06$ & 10.24 & 0.218561 & \phantom{$-$}84.884209 & $24.39 \pm 0.06 $ \\
2747550219666016765 & 274.755022 & 66.601677 & $22.08 \pm 0.06$ & \phantom{0}7.13 & 0.272215 & \phantom{$-$}26.869389 & $24.21 \pm 0.06$ \\
\hline
\end{tabular}}
\tablefoot{
\tablefoottext{$\dagger$}{Table\,\ref{appendix:struct-param1} is only available in electronic form at the CDS via anonymous ftp to cdsarc.u-strasbg.fr (130.79.128.5) or via \url{http://cdsweb.u-strasbg.fr/cgi-bin/qcat?J/A+A/}.}
}
\smallskip
\end{table*}

\begin{sidewaystable}[hp!]
\centering
\newcommand{\pd}{\phantom{1}}
\setlength{\tabcolsep}{3.25pt}
\caption{Aperture magnitudes and extinction corrections (EC) in the \Euclid \IE, \YE, \JE, and \HE\ bands for each dwarf galaxy candidate.\tablefootmark{$\dagger$}}
\label{appendix:struct-param2}
\smallskip
\smallskip
\resizebox{\textwidth}{!}{
\begin{tabular}{ccccccccc}
\hline
\hline
&&&&&&&\\[-7pt]
\omit\hfil ID \hfil & \omit\hfil $I_{{\rm E,2FWHM}}$ \hfil & \omit\hfil EC (\IE) \hfil & \omit\hfil $Y_{{\rm E,2FWHM}}$ \hfil & \omit\hfil EC (\YE) \hfil & \omit\hfil $J_{{\rm E,2FWHM}}$ \hfil & \omit\hfil EC (\JE) \hfil & \omit\hfil $H_{{\rm E,2FWHM}}$ \hfil & \omit\hfil EC (\HE) \hfil \\
\omit\hfil  & \omit\hfil [mag] \hfil & \omit\hfil [mag] \hfil & \omit\hfil [mag] \hfil & \omit\hfil [mag] \hfil & \omit\hfil [mag] \hfil & \omit\hfil [mag] \hfil & \omit\hfil [mag] \hfil & \omit\hfil [mag] \hfil \\
&&&&&&&\\[-8pt]
\hline
&&&&&&&\\[-8pt]
2665542189652466511 & $21.88 \pm 0.07$ & $0.116 \pm 0.003$ & $21.45 \pm 0.06$ & $0.058 \pm 0.001$ & $21.36 \pm 0.05$ & $0.0398 \pm 0.0009$ & $21.27 \pm 0.04$ & $0.0258 \pm 0.0006$ \\
2665815442648786462 & $22.67 \pm 0.07$ & $0.103 \pm 0.002$ & $22.08 \pm 0.06$ & $0.052 \pm 0.001$ & $22.00 \pm 0.05$ & $0.0352 \pm 0.0007$ & $21.84 \pm 0.04$ & $0.0228 \pm 0.0005$ \\
2665837571652492797 & $22.13 \pm 0.06$ & $0.117 \pm 0.002$ & $21.69 \pm 0.06$ & $0.059 \pm 0.001$ & $21.59 \pm 0.05$ & $0.0399 \pm 0.0007$ & $21.50 \pm 0.04$ & $0.0258 \pm 0.0004$ \\
2665847551651610342 & $20.69 \pm 0.07$ & $0.112 \pm 0.002$ & $20.32 \pm 0.05$ & $0.056 \pm 0.001$ & $20.20 \pm 0.04$ & $0.0384 \pm 0.0007$ & $20.11 \pm 0.03$ & $0.0248 \pm 0.0005$ \\
2665914298650060584 & $21.75 \pm 0.08$ & $0.110 \pm 0.003$ & $21.33 \pm 0.06$ & $0.055 \pm 0.001$ & $21.18 \pm 0.07$ & $0.038 \pm 0.001$ & $21.07 \pm 0.04$ & $0.0243 \pm 0.0006$ \\
\vdots & \vdots & \vdots & \vdots & \vdots & \vdots & \vdots & \vdots &\vdots \\
2747385300663221364 & $22.80 \pm 0.07$ & $0.125 \pm 0.002$ & $22.42 \pm 0.08$ & $0.063 \pm 0.001$ & $22.24 \pm 0.06$ & $0.0426 \pm 0.0007$ & $22.27 \pm 0.06$ & $0.0276 \pm 0.0005$ \\
2747397180663791709 & $20.42 \pm 0.06$ & $0.122 \pm 0.002$ & $19.79 \pm 0.04$ & $0.0614 \pm 0.0009$ & $19.72 \pm 0.04$ & $0.0418 \pm 0.0006$ & $19.63 \pm 0.03$ & $0.0271 \pm 0.0004$ \\
2747513424663839120 & $21.41 \pm 0.06$ & $0.122 \pm 0.002$ & $20.93 \pm 0.05$ & $0.0614 \pm 0.0009$ & $20.82 \pm 0.04$ & $0.0418 \pm 0.0006$ & $20.68 \pm 0.03$ & $0.0271 \pm 0.0004$ \\
2747532158662611810 & $22.49 \pm 0.06$ & $0.125 \pm 0.002$ & $22.18 \pm 0.07$ & $0.0627 \pm 0.0009$ & $22.08 \pm 0.06$ & $0.0427 \pm 0.0006$ & $22.11 \pm 0.06$ & $0.0276 \pm 0.0004$ \\
2747550219666016765 & $22.51 \pm 0.06$ & $0.125 \pm 0.002$ & $22.16 \pm 0.07$ & $0.0627 \pm 0.0009$ & $22.01 \pm 0.05$ & $0.0427 \pm 0.0006$ & $21.90 \pm 0.04$ & $0.0276 \pm 0.0004$ \\
\hline
\end{tabular}}
\tablefoot{
\tablefoottext{*}{Denotes dwarf candidates for which the colour information is affected by image defects.}\\
\tablefoottext{$\dagger$}{Table\,\ref{appendix:struct-param2} is only available in electronic form at the CDS via anonymous ftp to cdsarc.u-strasbg.fr (130.79.128.5) or via \url{http://cdsweb.u-strasbg.fr/cgi-bin/qcat?J/A+A/}.}
}
\smallskip
\end{sidewaystable}

\begin{table*}[ht!]
\caption{Stellar mass estimates of the dwarf galaxy candidates.\tablefootmark{$\dagger$}}
\label{appendix:mstar1}

\newcommand\Topspace{\rule{0pt}{2.3ex}}
\newcommand\Botspace{\rule[-1.3ex]{0pt}{0pt}}

\centering
{\renewcommand{\arraystretch}{1.4}
\begin{tabular}{ccccc}
\hline
\hline
\noalign{\smallskip}
\omit\hfil ID \hfil & \omit\hfil RA \hfil & \omit\hfil Dec \hfil & \omit\hfil $M_{\ast,z=0.1}$ \hfil & \omit\hfil $M_{\ast,z=0.05}$ \hfil \\
\omit\hfil  & \omit\hfil [deg] \hfil & \omit\hfil [deg] \hfil & \omit\hfil [$\log_{10}{M_\ast / M_\odot}$] \hfil & \omit\hfil [$\log_{10}{M_\ast / M_\odot}$] \hfil \Topspace \Botspace \\
\hline 
\noalign{\smallskip}
2665542189652466511 & 266.554219 & 65.246651 & $8.49^{+0.15}_{-0.08}$ & $7.83^{+0.41}_{-0.14}$ \\
2665815442648786462 & 266.581544 & 64.878646 & $8.50^{+0.20}_{-0.20}$ & $7.90^{+0.20}_{-0.20}$ \\
2665837571652492797 & 266.583757 & 65.249280 & $8.30^{+0.10}_{-0.10}$ & $7.61^{+0.08}_{-0.20}$ \\
2665847551651610342 & 266.584755 & 65.161034 & $8.87^{+0.30}_{-0.03}$ & $8.30^{+0.04}_{-0.04}$ \\
2665914298650060584 & 266.591430 & 65.006058 & $8.51^{+0.20}_{-0.05}$ & $8.00^{+0.20}_{-0.20}$ \\
\vdots & \vdots & \vdots & \vdots & \vdots \\
2747385300663221364 & 274.738530 & 66.322136 & $7.96^{+0.20}_{-0.13}$ & $7.30^{+0.20}_{-0.20}$ \\
2747397180663791709 & 274.739718 & 66.379171 & $9.21^{+0.04}_{-0.04}$ & $8.65^{+0.05}_{-0.20}$ \\
2747513424663839120 & 274.751342 & 66.383912 & $8.71^{+0.07}_{-0.10}$ & $7.93^{+0.13}_{-0.20}$ \\
2747532158662611810 & 274.753216 & 66.261181 & $7.97^{+0.11}_{-0.20}$ & $6.72^{+0.12}_{-0.12}$ \\
2747550219666016765 & 274.755022 & 66.601677 & $8.00^{+0.20}_{-0.20}$ & $7.50^{+0.20}_{-0.20}$ \Botspace \\
\hline 
\end{tabular}
}
\tablefoot{
\tablefoottext{$\dagger$}{Table\,\ref{appendix:mstar1} is only available in electronic form at the CDS via anonymous ftp to cdsarc.u-strasbg.fr (130.79.128.5) or via \url{http://cdsweb.u-strasbg.fr/cgi-bin/qcat?J/A+A/}.}
}
\end{table*}

\begin{table*}[ht!]
\caption{Stellar mass estimates of the dwarf galaxy candidates calculated with different models.\tablefootmark{$\dagger$}}
\label{appendix:mstar2}

\newcommand\Topspace{\rule{0pt}{2.3ex}}
\newcommand\Botspace{\rule[-1.3ex]{0pt}{0pt}}

\centering
{\renewcommand{\arraystretch}{1.4}
\begin{tabular}{cccccc}
\hline
\hline
\noalign{\smallskip}
\omit\hfil ID \hfil & \omit\hfil RA \hfil & \omit\hfil Dec \hfil & \omit\hfil $M_{\ast,z=z_\mathrm{spec}}$ \hfil & \omit\hfil $M_{\ast,z=z_\mathrm{spec} (\mathrm{\Euclid\,bands})}$ \hfil & \omit\hfil $M_{\ast,\mathrm{z}=\mathrm{z}_\mathrm{spec} (\mathrm{single-burst})}$ \hfil \\
\omit\hfil  & \omit\hfil [deg] \hfil & \omit\hfil [deg] \hfil & \omit\hfil [$\log_{10}{M_\ast / M_\odot}$] \hfil & \omit\hfil [$\log_{10}{M_\ast / M_\odot}$] \hfil & \omit\hfil [$\log_{10}{M_\ast / M_\odot}$] \hfil \Topspace \Botspace \\
\hline
\noalign{\smallskip}
2668226817651115008 & 266.822682 & 65.111501 & $8.61^{+0.06}_{-0.11}$ & $8.92^{+0.09}_{-0.13}$ & $8.63^{+0.04}_{-0.04}$ \\
2668228983656635904 & 266.822898 & 65.663601 & $8.44^{+0.08}_{-0.09}$ & $8.60^{+0.20}_{-0.20}$ & $8.19^{+0.05}_{-0.05}$ \\
2668283271656967168 & 266.828327 & 65.696713 & $8.56^{+0.05}_{-0.07}$ & $8.00^{+0.30}_{-0.40}$ & $8.52^{+0.04}_{-0.04}$ \\
2668500153654553088 & 266.850015 & 65.455303 & $7.97^{+0.03}_{-0.03}$ & $8.00^{+0.20}_{-0.20}$ & $7.86^{+0.03}_{-0.03}$ \\
2668644262651000320 & 266.864426 & 65.100031 & $8.89^{+0.11}_{-0.08}$ & $9.35^{+0.14}_{-0.20}$ & $8.91^{+0.03}_{-0.03}$ \\
\vdots & \vdots & \vdots & \vdots & \vdots & \vdots\\
2730261537666017792 & 273.026154 & 66.601767 & $6.96^{+0.05}_{-0.04}$ & $9.53^{+0.20}_{-0.13}$ & $8.25^{+0.03}_{-0.03}$ \\
2732068271664329216 & 273.206827 & 66.432926 & $8.56^{+0.03}_{-0.04}$ & $8.64^{+0.15}_{-0.13}$ & $8.57^{+0.03}_{-0.03}$ \\
2732101421668001280 & 273.210142 & 66.800147 & $8.48^{+0.04}_{-0.04}$ & $8.00^{+0.20}_{-0.50}$ & $8.50^{+0.03}_{-0.03}$ \\
2733911470664385024 & 273.391147 & 66.438544 & $8.13^{+0.06}_{-0.13}$ & $8.70^{+0.10}_{-0.20}$ & $8.01^{+0.06}_{-0.06}$ \\
2735409222663873024 & 273.540922 & 66.387278 & $8.86^{+0.04}_{-0.04}$ & $9.17^{+0.11}_{-0.20}$ & $8.87^{+0.03}_{-0.03}$ \Botspace \\
\hline 
\end{tabular}
}
\tablefoot{
\tablefoottext{$\dagger$}{Table\,\ref{appendix:mstar2} is only available in electronic form at the CDS via anonymous ftp to cdsarc.u-strasbg.fr (130.79.128.5) or via \url{http://cdsweb.u-strasbg.fr/cgi-bin/qcat?J/A+A/}.}
}
\end{table*}

\clearpage

\end{appendix}

\end{document}